\documentstyle[aps,preprint,eqsecnum]{revtex}
\begin{document}
\draft
 
\tighten
{\tighten

\title{Quantum Monte Carlo calculations of $A=8$ nuclei}
\author{R.\ B.\ Wiringa\cite{rbw} and Steven C.\ Pieper\cite{scp}}
\address{Physics Division, Argonne National Laboratory, 
         Argonne, IL 60439}
\author{J.\ Carlson\cite{jc}}
\address{Theoretical Division, Los Alamos National Laboratory, Los Alamos
         New Mexico 87545}
\author{V. R. Pandharipande\cite{vrp}}
\address{Physics Department, University of Illinois at Urbana-Champaign,
         1110 West Green St., Urbana, Illinois 61801}
\date{\today}
\maketitle

\begin{abstract}
We report quantum Monte Carlo calculations of ground and low-lying excited
states for $A=8$ nuclei using a realistic Hamiltonian containing the
Argonne $v_{18}$ two-nucleon and Urbana IX three-nucleon potentials.
The calculations begin with correlated eight-body wave functions that
have a filled $\alpha$-like core and four p-shell nucleons $LS$-coupled to 
the appropriate $(J^{\pi};T)$ quantum numbers for the state of interest.
After optimization, these variational wave functions are used as input
to a Green's function Monte Carlo calculation made with a new constrained 
path algorithm.
We find that the Hamiltonian produces a $^8$Be ground state that is within 
2 MeV of the experimental resonance, but the other eight-body energies are
progressively worse as the neutron-proton asymmetry increases.
The $^8$Li ground state is stable against breakup into subclusters, but the
$^8$He ground state is not. 
The excited state spectra are in fair agreement with experiment, with both
the single-particle behavior of $^8$He and $^8$Li and the collective rotational
behavior of $^8$Be being reproduced.
We also examine energy differences in the $T=1$ and 2 isomultiplets and 
isospin-mixing matrix elements in the excited states of $^8$Be.
Finally, we present densities, momentum distributions, and studies of the 
intrinsic shapes of these nuclei, with $^8$Be exhibiting a definite 
$2\alpha$ cluster structure.
\end{abstract}
 

}
 
\narrowtext

\section{INTRODUCTION}

The $A=8$ nuclei have many interesting properties that we would like to
understand on the basis of the bare interactions between individual nucleons.
The three strong-stable nuclei, $^8$He, $^8$Li, and $^8$B, decay by weak
processes with half-lives of $\alt 1$ second down to $^8$Be, which then 
immediately fissions to two $^4$He nuclei; thus there are no long-lived $A=8$ nuclei.
These facts have tremendous consequences for the nature of the universe we 
live in: they make the production of elements beyond $A=7$ in the early universe 
very difficult and help give stars like the sun a long stable lifetime.
The ground and excited states of the $A=8$ nuclei also provide a very sensitive 
testing ground for models of nuclear forces. 
In particular, $^8$He is the most neutron-rich strong-stable nucleus, and thus an
ideal place to study the isospin dependence of the three-nucleon force.
The $T=1,2$ isomultiplets are also a good place to look at charge-independence 
breaking, while some excited states in $^8$Be display significant isospin 
mixing.

Previously we have reported variational Monte Carlo (VMC) and Green's function
Monte Carlo (GFMC) calculations for the $A=6,7$ nuclei~\cite{PPCPW97}.
In this paper we extend these calculations to the ground and excited states
of $A=8$ nuclei.
(Some preliminary $A=8$ results were given in a number of conference
proceedings~\cite{Confs97_98}.)
We first construct a trial function, $\Psi_T$, with the proper
$(J^{\pi};T)$ quantum numbers and antisymmetry, and optimize the energy
expectation value. 
We then use $\Psi_T$ as the starting point for a GFMC calculation, which 
projects out the exact lowest-energy state by the Euclidean propagation
$\Psi_0 = \lim_{\tau \rightarrow \infty} \exp [ - ( H - E_0) \tau ] \Psi_T$.
We believe the resulting energy estimates are accurate to 1 to 2\% of
the binding energy in most cases.

In our previous work, the Argonne $v_{18}$ plus Urbana IX Hamiltonian 
was fairly successful in generating the excitation spectra of the $A=6,7$ 
nuclei, but did not give quite enough binding in the lithium isotopes, 
while $^6$He was unstable against breakup.
One of our main goals in studying the $A=8$ nuclei is to continue to test
the Hamiltonian, and lay the ground work for studies of more sophisticated
and accurate three-nucleon forces.
The Hamiltonian is reviewed in Sec.\ II.
The variational wave functions and calculations are described in Sec.\ III.
Our GFMC calculations make use of a new, more efficient algorithm 
called constrained-path propagation, which is described in Sec.\ IV.
Results for the ground-state energies, excitation spectra, isomultiplet
differences, and isospin-mixing matrix elements are given in Sec.\ V.
In Sec.\ VI we present various density distributions, and in Sec.\ VII we 
discuss the intrinsic shapes of these nuclei.
Our conclusions are given in Sec.\ VIII.

\section{HAMILTONIAN}

These calculations all use the same realistic Hamiltonian, which includes 
a nonrelativistic one-body kinetic energy, the Argonne $v_{18}$ 
two-nucleon potential~\cite{WSS95} and the Urbana IX three-nucleon 
potential~\cite{PPCW95}:
\begin{equation}
   H = \sum_{i} K_{i} + \sum_{i<j} v_{ij} + \sum_{i<j<k} V_{ijk} \ .
\end{equation}
The kinetic energy operator has charge-independent (CI) and
charge-symmetry-breaking (CSB) components, the latter due to
the difference in proton and neutron masses,
\begin{equation}
   K_{i} = K^{CI}_{i} + K^{CSB}_{i} \equiv -\frac{\hbar^2}{4}
     (\frac{1}{m_{p}} + \frac{1}{m_{n}}) \nabla^{2}_{i}
    -\frac{\hbar^2}{4} (\frac{1}{m_{p}} - \frac{1}{m_{n}})\tau_{zi}
     \nabla^{2}_{i} \ .
\end{equation}

The Argonne $v_{18}$ model~\cite{WSS95} is one of a class of new, highly
accurate $N\!N$ potentials that fit both $pp$ and $np$ scattering data up to
350 MeV with a $\chi^2/$datum near 1.
The potential can be written as a sum of electromagnetic and
one-pion-exchange terms and a shorter-range phenomenological part:
\begin{equation}
   v_{ij} = v^{\gamma}_{ij} + v^{\pi}_{ij} + v^{R}_{ij} \ .
\end{equation}
The electromagnetic terms include one- and two-photon-exchange Coulomb
interactions, vacuum polarization, Darwin-Foldy, and magnetic moment terms,
with appropriate proton and neutron form factors.
The one-pion-exchange part of the potential includes the charge-dependent (CD)
terms due to the difference in neutral and charged pion masses.
The shorter-range part has about 40 parameters which are adjusted to
fit the $pp$ and $np$ scattering data, the deuteron binding energy, and
also the $nn$ scattering length.
The one-pion-exchange and the remaining phenomenological part of the potential
can be written as a sum of eighteen operators,
\begin{equation}
       v^{\pi}_{ij} + v^{R}_{ij} = \sum_{p=1,18} v_{p}(r_{ij}) O^{p}_{ij} \ .
\label{eq:operator}
\end{equation}
The first fourteen are charge-independent,
\begin{eqnarray}
O^{p=1,14}_{ij} = [1, {\bf\sigma}_{i}\cdot{\bf\sigma}_{j}, S_{ij},
{\bf L\cdot S},{\bf L}^{2},{\bf L}^{2}{\bf\sigma}_{i}\cdot{\bf\sigma}_{j},
({\bf L\cdot S})^{2}]\otimes[1,{\bf\tau}_{i}\cdot{\bf\tau}_{j}] \ ,
\end{eqnarray}
and the last four,
\begin{equation}
   O^{p=15,18}_{ij} = [1, {\bf\sigma}_{i}\cdot{\bf\sigma}_{j},
S_{ij}]\otimes T_{ij} , \tau_{zi}+\tau_{zj} \ ,
\end{equation}
break charge independence.
We will refer to the potential from the $p=15-17$ terms as $v^{CD}$ and from
the $p=18$ term as $v^{CSB}$.
We note that in the context of isospin symmetry the CI, CSB
and CD terms are respectively isoscalar, isovector and isotensor.

The two-nucleon potential is supplemented by a three-nucleon interaction from 
the Urbana series of models~\cite{CPW83}, including both long-range two-pion 
exchange and a short-range phenomenological component:
\begin{equation}
   V_{ijk} = V^{2\pi}_{ijk} + V^{R}_{ijk} \ .
\end{equation}
The two-pion-exchange term can be expressed simply as a sum of anticommutator
and commutator terms,
\begin{equation}
   V^{2\pi}_{ijk} = \sum_{cyclic} V^{2\pi,A}_{ij;k} + V^{2\pi,C}_{ij;k} \ .
\end{equation}
Here
\begin{equation}
   V^{2\pi,A}_{ij;k} = A_{2\pi} \{X^{\pi}_{ik},X^{\pi}_{jk}\} 
                       \{\tau_{i}\cdot\tau_{k},\tau_{j}\cdot\tau_{k}\} \ , 
\label{eq:2piA}
\end{equation}
and
\begin{equation}
   V^{2\pi,C}_{ij;k} = \frac{1}{4}A_{2\pi}  [X^{\pi}_{ik},X^{\pi}_{jk}] 
                        [\tau_{i}\cdot\tau_{k},\tau_{j}\cdot\tau_{k}] \ ,
\end{equation}
with $X^{\pi}_{ij} = Y(r_{ij})\sigma_{i}\cdot\sigma_{j} + T(r_{ij})S_{ij}$
as the basic one-pion exchange operator.  The $V^{R}_{ijk}$ has no spin-isospin 
dependence, and in 
the Urbana model IX~\cite{PPCW95} its strength and $A_{2\pi}$ are 
adjusted to reproduce the binding 
energy of $^3$H and give a reasonable saturation density in nuclear matter 
when used with Argonne $v_{18}$.

The CD and CSB terms in $H$ are fairly weak, so we can treat them
conveniently as a first-order perturbation and use a wave function 
of good isospin, which is significantly more compact.
Also, direct GFMC calculations with the spin-dependent terms that 
involve the square of the momentum operator can have large statistical
fluctuations, as discussed in Ref.~\cite{PPCPW97}.
Thus we construct the GFMC propagator with a simpler isoscalar
Hamiltonian,
\begin{equation}
     H^{\prime} = \sum_{i}K^{CI}_{i} + \sum_{i<j} v^{\prime}_{ij}
                + \sum_{i<j<k}V^{\prime}_{ijk} \ ,
\label{eq:hgfmc}
\end{equation}
where $v^{\prime}_{ij}$ is defined as
\begin{equation}
     v^{\prime}_{ij}=\sum_{p=1,8} v^{\prime}_{p}(r_{ij}) O^{p}_{ij}
                                + v^{\prime}_{C}(r_{ij}) \ .
\label{eq:vprime}
\end{equation}
The interaction $v^{\prime}_{ij}$ has only eight operator terms, with operators
$[1, {\bf\sigma_{i}}\cdot{\bf\sigma_{j}}, S_{ij},{\bf L\cdot S}]
\otimes[1,{\bf\tau_{i}}\cdot{\bf\tau_{j}}]$, chosen such that it equals the 
isoscalar part of the full interaction in all $S$ and $P$ waves as well as in 
the $^{3}D_{1}$ wave and its coupling to the $^{3}S_{1}$.
The isoscalar part of the $pp$ Coulomb interaction, $v^{\prime}_{C}$,
is also included in $H^{\prime}$.
Detailed expressions are given in Ref.~\cite{PPCPW97}.
The $v^{\prime}_{ij}$ is a little more attractive than $v_{ij}$, so we 
compensate by using a $V^{\prime}_{ijk}$ that is adjusted to keep
$\langle H^{\prime} \rangle \approx \langle H \rangle$; this should help
prevent the GFMC propagation from producing excessively large densities 
due to overbinding.
The small contribution of $(H-H^{\prime})$ is calculated perturbatively.

\section{VARIATIONAL MONTE CARLO}

The variational method can be used to obtain approximate solutions to 
the many-body Schr\"{o}dinger equation, $H\Psi = E\Psi$, for a wide 
range of nuclear systems, including few-body nuclei, light closed-shell 
nuclei, nuclear matter, and neutron stars~\cite{W93}.
A suitably parameterized wave function, $\Psi_V$, is used to calculate 
an upper bound to the exact ground-state energy,
\begin{equation}
   E_V = \frac{\langle \Psi_V | H | \Psi_V \rangle}
              {\langle \Psi_V   |   \Psi_V \rangle} \geq E_0 \ .
\label{eq:expect}
\end{equation}
The parameters in $\Psi_V$ are varied to minimize $E_V$, and the lowest 
value is taken as the approximate ground-state energy. 

Upper bounds to excited states are also obtainable, either
from standard VMC calculations if they have different quantum
numbers from the ground state, or from small-basis diagonalizations
if they have the same quantum numbers.
The corresponding $\Psi_V$ can then be used to calculate other 
properties, such as electromagnetic form factors~\cite{WS98} and
spectroscopic factors~\cite{LWW99}, or it can be used as the starting 
point for a Green's function Monte Carlo calculation.
In this section we first describe our {\it ansatz} for $\Psi_V$ for the 
$A=8$ nuclei and then briefly review how the expectation value, 
Eq.~(\ref{eq:expect}), is evaluated and the parameters of 
$\Psi_V$ are fixed.

\subsection{Wave Function}

Our best variational wave function for the nuclei studied here has the 
form~\cite{APW95}
\begin{equation}
     |\Psi_V\rangle = \Big[1 + \sum_{i<j<k}(U_{ijk}+U^{TNI}_{ijk}) 
                              + \sum_{i<j}U^{LS}_{ij} \Big]
                      |\Psi_P\rangle \ ,
\label{eq:bestpsiv}
\end{equation}
where the pair wave function, $\Psi_P$, is given by
\begin{equation}
     |\Psi_P\rangle = {\cal S}\prod_{i<j}(1+U_{ij})
                      |\Psi_J\rangle \ .
\label{eq:psip}
\end{equation}
The $U_{ij}$, $U^{LS}_{ij}$, $U_{ijk}$, and $U^{TNI}_{ijk}$ are 
noncommuting two- and three-nucleon correlation operators, and the 
${\cal S}$ is a symmetrization operator.
The $U_{ij}$ includes spin, isospin, and tensor operators, while 
the $U^{LS}_{ij}$ has spin-orbit operators, reflecting the operator
structure of the two-nucleon interaction, Eq.~(\ref{eq:operator}).
The $U_{ijk}$ is a nontrivial operator in spin-isospin space also induced 
by $v_{ij}$, while the $U^{TNI}_{ijk}$ reflects the structure of the
three-nucleon interaction.
All these correlations are discussed fully in Refs.~\cite{PPCPW97,APW95,W91}.
The two-body correlations are generated by the solution of coupled 
differential equations with embedded variational parameters. 
We have found that the parameters optimized for the $\alpha$-particle are 
near optimal for use in the light p-shell nuclei.
Likewise, the best parameters for the three-body correlations are
remarkably constant for different s- and p-shell nuclei, so they have
not been changed from the previous work.

The form of the totally antisymmetric Jastrow wave function, 
$\Psi_J$, depends on the nuclear state under investigation.
For s-shell nuclei we use the simple form
\begin{equation}
     |\Psi_J\rangle = \prod_{i<j<k}f^c_{ijk} ~
                      \prod_{i<j}f_c(r_{ij}) ~
                     |\Phi_A(JMTT_{3})\rangle \ .
\label{eq:jastrow}
\end{equation}
Here $f_c(r_{ij})$ and $f^c_{ijk}$ are central two- and three-body 
correlation functions and for the $\alpha$-particle,
\begin{equation}
   |\Phi_{4}(0 0 0 0) \rangle
     = {\cal A} |p\uparrow p\downarrow n\uparrow n\downarrow \rangle \ .
\end{equation}
The Jastrow wave function for the light p-shell nuclei is significantly
more complicated due to the requirements of antisymmetry.
Expressions for $A=6,7$ nuclei are given in Ref.~\cite{PPCPW97};
the present $\Psi_J$ is a straightforward extension.
We use $LS$ coupling to obtain the desired $JM$ value of a given state, 
as suggested by standard shell-model studies~\cite{CK65}.
We also need to specify the spatial symmetry $[n]$ of the angular
momentum coupling of four p-shell nucleons~\cite{BM69}.
Different possible $LS[n]$ combinations lead to multiple components in the
Jastrow wave function.
We allow for the possibility that the central correlations 
$f_{c}(r_{ij})$ could depend upon the shells ($s$ or $p$) occupied by the 
particles and on the $LS[n]$ coupling.
The Jastrow wave function is taken as
\begin{eqnarray}
  |\Psi_{J}\rangle &=& {\cal A} \left\{\right.
     \prod_{i<j<k \leq 4} f^{c}_{ijk} 
     \prod_{i<j \leq 4}f_{ss}(r_{ij})
     \prod_{k \leq 4} f_{sp}(r_{k5}) f_{sp}(r_{k6})
                      f_{sp}(r_{k7}) f_{sp}(r_{k8})  \nonumber\\
                   && \sum_{LS[n]}
     \Big( \beta_{LS[n]} \prod_{5 \leq l<m \leq 8} f^{[n]}_{pp}(r_{lm})
    |\Phi_{8}(LS[n]JMTT_{3})_{1234:5678}\rangle \Big) \left.\right\} \ .
\label{eq:jastrow8}
\end{eqnarray}
The operator ${\cal A}$ indicates an antisymmetric sum over all possible
partitions of the eight particles into 4 s-shell and 4 p-shell ones.
For the two-body correlations we use $f_{ss}(r)=f_c(r)$ from the 
$^{4}$He wave function, while
\begin{eqnarray}
   f_{sp}(r)       &=& f_c(r) + c_{sp}      (1-\exp[-(r/d_{sp})^2]) \ , \\
\end{eqnarray}
and
\begin{eqnarray}
   f^{[n]}_{pp}(r) &=& f_c(r) + c^{[n]}_{pp}(1-\exp[-(r/d_{pp})^2]) \ ,
\label{eq:flspp}
\end{eqnarray}
where we have supplemented the $f_c(r)$ with a long-range tail.
The $c_{sp}$, $d_{sp}$, etc., are variational parameters, whose values 
are given in Table~\ref{table:params8}.
For the three-body correlations, our best present trial function has 
the $f^{c}_{ijk}$ acting only within the s-shell.

The $LS[n]$ components of the single-particle wave function are given by:
\begin{eqnarray}
     |\Phi_{8}(LS[n]JMTT_{3})_{1234:5678}\rangle &=& 
     |\Phi_{4}(0 0 0 0)_{1234} 
     \phi^{LS}_{p}(R_{\alpha 5}) \phi^{LS}_{p}(R_{\alpha 6})
     \phi^{LS}_{p}(R_{\alpha 7}) \phi^{LS}_{p}(R_{\alpha 8})
     \nonumber \\
 &&  \left\{\right. [Y_{1m_l}(\Omega_{\alpha 5}) Y_{1m_l'}(\Omega_{\alpha 6})
         Y_{1m_l''}(\Omega_{\alpha 7}) Y_{1m_l'''}(\Omega_{\alpha 8})]_{LM_L}
         \nonumber \\
 &&  \left.\times [\chi_{5}(\case{1}{2}m_s) \chi_{6}(\case{1}{2}m_s')
         \chi_{7}(\case{1}{2}m_s'') \chi_{8}(\case{1}{2}m_s''')]_{SM_S}
         \right\}_{JM} \nonumber \\
 &&  \times [\nu_{5}(\case{1}{2}t_3) \nu_{6}(\case{1}{2}t_3')
         \nu_{7}(\case{1}{2}t_3'') \nu_{8}(\case{1}{2}t_3''')]_{TT_3}\rangle \ .
\end{eqnarray}
The $\phi^{LS}_{p}(R_{\alpha k})$ are $p$-wave solutions of a particle 
of reduced mass $\case{4}{5}m_N$ in an effective $\alpha$-$N$ potential:
\begin{equation}
   V_{\alpha N}(r) = V^{WS}_{\alpha N}(r) + V^C_{\alpha N}(r) \ .
\end{equation}
They are functions of the distance between the center of mass 
of the $\alpha$ core (which contains particles 1-4 in this partition)
and nucleon $k$, and again may be different for different $LS$ 
components.
For each state considered in the present work, we have used
bound-state asymptotic conditions for the $\phi^{LS}_{p}$, even if
the state is particle unstable.
The Woods-Saxon potential
\begin{equation}
   V^{WS}_{\alpha N}(r) = V^{LS}_p [1+exp(\frac{r-R_p}{a_p})]^{-1} \ ,
\label{eq:spwell}
\end{equation}
has variational parameters $V^{LS}_p$, $R_p$, and $a_p$ 
whose values are given in Table~\ref{table:params8}.
The Coulomb potential is obtained by folding over nuclear form factors:
\begin{eqnarray}
  V^C_{\alpha N}(r) = 2Q\frac{e^2}{r} \left\{\right. 
                 1 &-& \frac{1}{2}exp(-x_{\alpha})
            [2+x_{\alpha}+\frac{4}{1-y^2}] [1-y^{-2}]^{-2}  \nonumber\\
                   &-& \frac{1}{2}exp(-x_p)
            [2+x_p+\frac{4}{1-y^{-2}}] [1-y^2]^{-2} \left.\right\} \ .
\label{eq:vcan}
\end{eqnarray}
Here $x_{\alpha} = \sqrt{12}r/r_{\alpha}$, $x_p = \sqrt{12}r/r_p$, and
$y=r_{\alpha}/r_p$, with the charge radii $r_{\alpha}=1.65$ fm and
$r_p = 0.81$ fm.
This additional potential term can be used with strength $Q=0$, $\case{1}{4}$, 
or $\case{1}{2}$ for $^8$He, $^8$Li, or $^8$Be, respectively, corresponding 
to the average Coulomb interaction between the $\alpha$ core and a p-shell
nucleon.
The wave function is translationally invariant, hence there is no 
spurious center of mass motion.

The experimental spectra for $A=8$ nuclei~\cite{AS88} are shown in 
Fig.~\ref{fig:expt8}.
The ground state of $^8$He is strong stable, but decays by $\beta^-$ 
emission with a half life of 119 ms.
One excited state is identified at 3.59 $\pm$ 0.05 MeV~\cite{he8-2plus}, above the 
threshold for decay to $^6$He+2$n$.
In the shell model, the $(J^{\pi};T) = (0^+;2)$ ground state
is predominantly a $^{2S+1}L[n] = ^1$S[22] state, where we use 
spectroscopic notation to denote the total $L$ and $S$ and 
the Young pattern $[n]$ to indicate the spatial symmetry.
The $(2^+;2)$ first excited state is predominantly a $^1$D[22] state.
We also allow for possible $^3$P[211] admixtures, which are the only other 
available configurations in the p-shell, in our $\Psi_J$.
After other parameters in the trial function have been optimized, we 
make a series of calculations in which the $\beta_{LS[n]}$ may be 
different in the left- and right-hand-side wave functions to obtain the 
diagonal and off-diagonal matrix elements of the Hamiltonian and the 
corresponding normalizations and overlaps.
We diagonalize the resulting matrices to find the $\beta_{LS[n]}$ 
eigenvectors.
The shell-model wave functions are orthonormal, but the correlated 
$\Psi _V$ are not.  
Hence the diagonalizations use generalized eigenvalue routines including
overlap matrices.
We also calculate the position of the three predominantly $^3$P[211] 
states, with $(J^{\pi};T) = (2^+;2)$, $(1^+;2)$, and $(0^+;2)$;
none of these have been identified experimentally.
The normalized $\beta_{LS[n]}$ for these different states are given in
Table~\ref{table:beta82}.

The ground state of $^8$Li is a strong stable $(2^+;1)$ state that 
decays by $\beta^-$ emission with a half life of 838 ms, and is 
predominantly $^3$P[31] in character.
The $(1^+;1)$ first excited state at 0.98 MeV excitation is also strong 
stable with a 12 fs $\gamma$-decay, and is primarily a mix of $^{1,3}$P[31] 
configurations.
The $(3^+;1)$ second excited state at 2.26 MeV is just above the threshold 
for breakup into $^7$Li+$n$ and is fairly narrow with a width of 33 keV, 
decaying by both $\gamma$ and $n$ emission.
A number of higher states have been identified, most having fairly large
widths.
Two special cases are the $(4^+;1)$ stretched state at 6.53 MeV excitation
with a width of 35 keV, which can only come from the $^3$F[31] p-shell
configuration, and the $(0^+;2)$ isobaric analog of $^8$He that 
occurs at 10.82 MeV with a width less than 12 keV.
The possible p-shell components in the $T=1$ states include the
$^{1,3}$P[31], $^{1,3}$D[31], $^{1,3}$F[31], $^3$S[22], $^3$D[22], 
and $^{1,3,5}$P[211] configurations. 
We include all but the lowest-symmetry [211] components in our $\Psi_J$
and calculate all possible first, second and some third states of 
given $(J^{\pi};T)$ through the diagonalization procedure discussed above.
Table~\ref{table:beta81} gives a summary of the $\beta_{LS[n]}$ amplitudes.

The $(0^+;0)$ ground state of $^8$Be is 92 keV above the threshold for
breakup into two $\alpha$-particles, with a width of only 7 eV;
it is an almost pure $^1$S[4] configuration.
The first and second excited states are very broad: a $(2^+;0)$ $^1$D[4] 
state at 3.04 MeV and a $(4^+;0)$ $^1$G[4] state at 11.4 MeV --- the 
spacing of an almost rigid rotor.
Indeed, as discussed below, $^8$Be appears to have the intrinsic deformation 
of a $2\alpha$ molecule.
Higher in the spectrum is the famous pair of isospin-mixed $(2^+;0+1)$ states 
at 16.63 and 16.92 MeV which have widths of order 100 keV; the 
$T=1$ component is the isobaric analog of the $^8$Li ground state.
There are similar $(1^+;0+1)$ and $(3^+;0+1)$ pairs near 18 and 19 MeV
which have widths less than 300 keV.
Many additional states have been identified above 18 MeV, most of them with
large widths, up to the $(0^+;2)$ isobaric analog of $^8$He
at 27.49 MeV.
There are also some negative-parity states in this region, which we have
not attempted to calculate.
In constructing $\Psi_J$, we use all p-shell states of symmetry [4] and
[31], but neglect those of symmetry [22] and [211], as shown in 
Table~\ref{table:beta80}.

The known experimental spectrum for $^8$B is similar to $^8$Li, except that
with the extra Coulomb repulsion, the $(2^+;1)$ ground state is just barely 
strong stable, decaying by $\beta+$ emission with a half life of 770 ms.
The $(1^+;1)$ first excited state is seen as a narrow resonance in $^7$Be+$p$
scattering, and the $(3^+;1)$ second excited state is much broader than
its $^8$Li analog.
The only other observed state is the $(0^+;2)$ isobaric analog to $^8$He
at 10.62 MeV.
The ground state for $^8$C is unstable against several possible breakup
channels, having a width of $\sim$230 keV, but its mass excess is known 
within 20 keV.
We have calculated the energies of these states in $^8$B and $^8$C so we can
study the energy differences in the $T=1$ and 2 isomultiplets.

The full $A=8$ wave function is constructed by acting on the 
$\Psi_J$, Eq.~(\ref{eq:jastrow8}), with the $U_{ij}$, 
$U^{LS}_{ij}$, $U_{ijk}$, and $U^{TNI}_{ijk}$ correlations.
Because of the tensor and other correlations in $U_{ij}$, many additional
symmetry components, beyond the explicit p-shell states discussed above,
are built up in the wave function.
In principle, the $U_{ij}$ could be generalized to be different 
according to whether particles $i$ and $j$ are in the s- or 
p-shell, but this would require a larger sum over the different 
partitions and would increase the computational cost by an order of 
magnitude.

For input to the GFMC algorithm, it is more efficient to use the 
somewhat simplified trial function
\begin{eqnarray}
|\Psi_{T}\rangle =  {\cal S} \prod_{i<j} \Big( 1+U_{ij} +
\sum_{k \neq i,j}\tilde{U}^{TNI}_{ij;k}   \Big)
|\Psi_J\rangle \ ,
\label{eq:psitgfmc}
\end{eqnarray}
where $\tilde{U}^{TNI}_{ij;k}$ is a truncated three-nucleon interaction 
correlation based on the short-range $V^{R}_{ijk}$ term and on the 
anticommutator part of the two-pion exchange, $V^{2\pi,A}_{ij;k}$,
which can be reduced to operators that depend only on the spins and isospins
of nucleons $i$ and $j$.
Thus the sum over $k$ can be made, leaving a two-body spin-isospin
operator that can be combined with $U_{ij}$; the result is calculable
with only a little more effort than just $U_{ij}$ alone.
This trial function gets the bulk of the energy, as shown below, but
for about half the computational effort of the full $\Psi_{V}$.

\subsection{Energy Evaluation}

The energy expectation value of Eq.~(\ref{eq:expect}) is evaluated using 
Monte Carlo integration.
A detailed technical description of the methods used here can be found 
in Refs.~\cite{W91,CW91,P96}.
Monte Carlo sampling is done both in configuration space and in the 
discrete order of operators in the symmetrized product of the pair wave 
function by following a Metropolis random walk.
The expectation value for an operator $O$ with the full wave function 
$\Psi_{V}$is given by
\begin{equation}
  \langle O \rangle = \frac
  { \sum_{p,q} \int d{\bf R} 
    \left[ \Psi_{V,p}^{\dagger}({\bf R}) O \Psi_{V,q}({\bf R}) / 
           W_{pq}({\bf R}) \right] W_{pq}({\bf R}) }
  { \sum_{p,q} \int d{\bf R}
    \left[ \Psi_{V,p}^{\dagger}({\bf R})   \Psi_{V,q}({\bf R}) /
           W_{pq}({\bf R}) \right] W_{pq}({\bf R}) } \ ,
\label{eq:vmc:expectation}
\end{equation}
where we have introduced a probability distribution, $W_{pq}({\bf R})$,
based on the approximate wave function $\Psi_{P}$ of Eq.~(\ref{eq:psip}),
\begin{equation}
   W_{pq}({\bf R}) = | {\rm Re}( \langle \Psi_{P,p}^{\dagger}({\bf R})
                                  \Psi_{P,q}({\bf R}) \rangle ) | \ .
\label{eq:vmc:weight}
\end{equation}
The subscripts $p$ and $q$ specify the order of operators on the left and 
right hand side of the pair wave functions, while the integration runs over the 
particle coordinates ${\bf R}=({\bf r}_1,{\bf r}_2,\ldots,{\bf r}_A)$.
This probability distribution is much less expensive to compute than one
using the full wave function of Eq.~(\ref{eq:bestpsiv}) with its spin-orbit 
and operator-dependent three-body correlations, but the denominator
of Eq.~(\ref{eq:vmc:expectation}) is typically within 1-2\% of unity.
Expectation values have a statistical error which can be estimated by 
the standard deviation $\sigma$ in either gaussian approximation or by using
block averaging schemes.

Our wave functions are vectors of 
$2^A \times I(A,T)$ complex numbers, 
\begin{equation}
  \Psi({\bf R}) = \sum_{\alpha} \psi_{\alpha}({\bf R}) |\alpha\rangle \ ,
\label{eq:psivec}
\end{equation}
where the $\psi_{\alpha}({\bf R})$ are the coefficients of each possible
spin-isospin state $|\alpha\rangle$ with specific third components of the
spins of each nucleon and the desired total isospin.
The CD and CSB force components are sufficiently small that we do not
worry about isospin mixing in the wave function, and the expression for the 
number of isospin states, $I(A,T)$, is given in Ref.~\cite{PPCPW97}.
This gives arrays with 3584, 5120, and 7168 elements for $^8$Be, $^8$He, 
and $^8$Li, respectively.
The spin, isospin, and tensor operators $O^{p=2,6}_{ij}$ contained in
the two-body correlation operator $U_{ij}$, and in the Hamiltonian are
sparse matrices in this basis.

Expectation values of the kinetic energy and spin-orbit potential 
require the computation of first derivatives and diagonal second 
derivatives of the wave function.
These are obtained by evaluating the wave function at $6A$ slightly 
shifted positions of the coordinates ${\bf R}$ and taking finite 
differences, as discussed in Ref.~\cite{W91}.
Potential terms quadratic in {\bf L} require mixed second derivatives, 
which can be obtained by additional wave function evaluations and 
finite differences.
A rotation trick can be used to reduce the number of additional 
locations at which the wave function must be evaluated~\cite{SPF89}.

In addition to calculating energies, we evaluate $\langle J^2 \rangle$
and $\langle J_z \rangle$ expectation values to verify that our wave
functions truly have the specified quantum numbers.
Another check is made on the antisymmetry of the Jastrow wave function 
by evaluating, at an initial randomized position,
\begin{equation}
  \frac{
  \Psi_J^{\dagger} [ 1 + P^x_{ij}P^{\sigma}_{ij}P^{\tau}_{ij} ] \Psi_J }
  { \Psi_J^{\dagger} \Psi_J } \ , \nonumber
\end{equation}
where $P^{x,\sigma,\tau}_{ij}$ are the space, spin, and isospin 
exchange operators.
This value should be exactly zero for an antisymmetric wave 
function, and it is in fact less than $10^{-9}$ for each pair of 
particles in each nuclear state that we study.

A major problem arises in minimizing the variational energy for 
p-shell nuclei using the above wave functions: there is no variational
minimum that gives reasonable rms radii.
For example, the variational energy for $^6$Li is slightly more bound
than for $^4$He, but is not more bound than for separated $^4$He and $^2$H
nuclei, so the wave function is not stable against breakup into $\alpha + d$
subclusters.
Consequently, the energy can be lowered toward the sum of $^4$He and $^2$H
energies by making the wave function more and more diffuse.
Such a diffuse wave function would not be useful for computing other 
nuclear properties, or as a starting point for the GFMC calculation.

On the basis of our work in $A$ = 6--7 nuclei, we believe that part 
of the problem is a fault of the Hamiltonian, but most of it is due to the
presence of small admixtures of highly excited states in the trial
function~\cite{PPCPW97}.
In that work, we constrained our search for optimal variational 
parameters by requiring the resulting point proton rms radius, $r_p$, 
to be close to the experimental values for $^6$Li and $^7$Li ground 
states.
Then we allowed only small variations in the construction of the $^6$He 
and $^7$Be ground states and all the excited or resonant states, 
for which there are no experimental measurements of the charge radii. 

In the present work, we have the additional complication that all the
$A=8$ nuclei are sufficiently short-lived that precise experimental
determinations of their radii exist do not exist.
Consequently, the variational parameters in Table~\ref{table:params8}
are chosen very close to those of our previous work, with only
a systematic reduction in the depth of the Woods-Saxon potential well
as $L$ increases, and an increase in the tail of the $f^{[n]}_{pp}(r)$
correlation as the spatial symmetry declines, as shown in 
Table~\ref{table:params8}.
These gradual changes help to insure that the radii of excited states
increase as the excitation energy increases.

The last step is always the diagonalization to determine the $\beta_{LS[n]}$ 
mixing coefficients of Tables~\ref{table:beta82}-\ref{table:beta80}.
Shell model lore tells us that the lowest state of any given 
$(J^{\pi};T)$ will be the state with maximal spatial symmetry and 
smallest $L$ that can be formed from the allowed couplings, e.g., 
the $^1$S[4] ground state in $^8$Be or the $^3$P[31] ground state in 
$^8$Li.
For the purposes of obtaining a variational upper bound and a GFMC 
starting point, we could settle for a $\Psi_V$ constructed using only 
that $LS[n]$ component.
However, by using more components, we can gain a significant amount of 
energy in some cases and this gain persists in our GFMC propagations.
For the $^8$He ground state, when the dominant $^1$S[22] piece is
supplemented by the $^3$P[211] term, the energy gain is 1.2 MeV.
In the case of $^8$Li ground state, where the dominant term is $^3$P[31],
addition of the three other contributing [31] symmetry states lowers the
energy 0.8 MeV; addition of the one [22] state (the next highest spatial
symmetry) gives only an additional 0.1 MeV. 
However, for the $^8$Be ground state, with the dominant $^1$S[4] term, 
addition of the one [31] symmetry piece gives only 0.1 MeV.
For this reason, we feel reasonably confident in truncating the $^8$Li
and $^8$Be p-shell bases after the top two symmetry states.

\section{GREEN'S FUNCTION MONTE CARLO}

A detailed description of the nuclear GFMC method, and many tests of its 
accuracy, are given in Ref.~\cite{PPCPW97}.  In this section we present 
a brief review of the method and then describe two improvements
that have been made since that publication.  In most of this section,
we will not make the distinction between $H^{\prime}$ and $H$; the reader
should remember that in fact we use the simpler $H^{\prime}$ in
our GFMC propagator and evaluate $\langle H -  H^{\prime} \rangle$
perturbatively.

\subsection{Review}

The GFMC method starts with the trial wave function, $\Psi_{T}$ of
Eq.~(\ref{eq:psitgfmc}), and projects out of it the exact lowest energy
state with the same quantum numbers, $\Psi_{0}$:
\begin{equation}
\Psi_0 = \lim_{\tau \rightarrow \infty} \Psi(\tau) \ , 
\end{equation}
\begin{equation}
\Psi(\tau) = e^{ - ( H - E_0) \tau } \Psi_T 
           = \left[e^{-({H}-E_{0})\triangle\tau}\right]^{n} \Psi_{T} \ .
\end{equation}
Here we have sliced the imaginary propagation time, $\tau$, into a 
number of small time steps, $\triangle\tau=\tau/n$.
The small-time-step Green's function, 
$G_{\alpha\beta}({\bf R},{\bf R}^{\prime})$, is a matrix function of 
$\bf R$ and ${\bf R}^{\prime}$ in spin-isospin space, with matrix elements
defined as
\begin{eqnarray}
G_{\alpha\beta}({\bf R},{\bf R}^{\prime})= \langle {\bf 
R},\alpha|e^{-({H}-E_{0})\triangle\tau}|{\bf R}^{\prime},\beta\rangle \ .
\label{eq:gfunction}
\end{eqnarray}
Then $\Psi({\bf R}_{n},\tau)$ is given by 
\begin{eqnarray}
\Psi({\bf R}_{n},\tau) = \int G({\bf R}_{n},{\bf R}_{n-1}) \cdots 
G({\bf R}_{1},{\bf R}_{0})\Psi_{T}({\bf R}_{0}) 
d{\bf R}_{n-1} \cdots d{\bf R}_{1}d{\bf R}_{0} \ .
\label{eq:gfmcpsi}
\end{eqnarray}

The small-time-step propagator used in Ref.~\cite{PPCPW97} is
\begin{eqnarray}
G_{\alpha\beta}({\bf R},{\bf R}^{\prime})& = &e^{E_0 \triangle\tau}
G_{0}({\bf R},{\bf R}^{\prime})
\exp\left[{- \frac{\triangle\tau}{2} 
\sum (V^{R}_{ijk}({\bf R})+ V^{R}_{ijk}({\bf R^{\prime}}))} \right]     \nonumber \\ 
& & \times \langle\alpha|I_{3}({\bf R})|\gamma\rangle\langle\gamma|
\left[{\cal S}\prod_{i<j}\frac{g_{ij}({\bf r}_{ij},{\bf r}_{ij}^{\prime})}
{g_{0,ij}({\bf r}_{ij},{\bf r}_{ij}^{\prime})} \right] |\delta\rangle
\langle\delta|I_{3}({\bf R}^{\prime})|\beta\rangle \ ,          
\label{eq:fullprop}
\end{eqnarray}
where $g_{ij}({\bf r}_{ij},{\bf r}_{ij}^{\prime})$ is the exact two-body
propagator, $g_{0,ij}({\bf r}_{ij},{\bf r}_{ij}^{\prime})$ is the free
two-body propagator, $G_{0}({\bf R},{\bf R}^{\prime})$ is the free
many-body propagator, $\alpha,\beta,\gamma,\delta$ are spin-isospin state
indices, and summation over $\gamma,\delta$ is implied.  
There is also an implicit sampling of the order of pairs in the symmetrized
product of Eq.~(\ref{eq:fullprop}).
The construction of the exact two-body propagator is described in 
Ref.~\cite{PPCPW97}.
The influence of the three-nucleon potential on the many-body propagator 
is broken into two pieces: the scalar $V^{R}_{ijk}$ which is easily 
exponentiated, and the $V^{2\pi}_{ijk}$ which is a more complicated operator 
in spin-isospin space.   The simplest treatment of this term in the 
TNI is to expand to first order in $\triangle\tau$,
\begin{equation}
I_{3}({\bf R}) = 1 - 
\frac{\triangle\tau}{2}\sum V^{2\pi}_{ijk}({\bf R}) \ ;
\end{equation}
in fact we use a more efficient procedure described below in subsection E.

The integrals in Eq.~(\ref{eq:gfmcpsi}) are evaluated stochastically by 
averaging over a set of $n$-step paths, 
${\bf P}_{n} = {\bf R}_{0},{\bf R}_{1},\cdots,{\bf R}_{n}$.  
The paths are chosen by first sampling a set of positions, ${\bf R}_{0}$, 
using a probability function based on $\Psi_T({\bf R}_{0})$, and then 
sequentially sampling the free Green's functions, 
$G_{0}({\bf R}_{i+1},{\bf R}_i)$, to generate ${\bf R}_{i+1}$ from ${\bf R}_i$.
We thus obtain a spin-isospin vector $\Psi({\bf P}_n)$ which is
one sample of the integrand of Eq.~(\ref{eq:gfmcpsi}).
A GFMC ``configuration'' consists of the position, ${\bf R}_n$, and
the vector $\Psi({\bf P}_n)$.
Branching and importance sampling, described in detail in Ref.~\cite{PPCPW97}, 
are used to obtain samples with probability proportional to the scalar
importance function $I$:
\begin{equation}
I [ \Psi({\bf P}_n), \Psi_{T,p}({\bf R}_n) ] = 
| {\rm Re} [ \sum_\alpha \psi_{\alpha}({\bf P}_n)^\dagger \psi_{T,p,\alpha}({\bf R}_n) ] | \\
+ \epsilon \sum_\alpha |  [  \psi_{\alpha}({\bf P}_n)^\dagger \psi_{T,p,\alpha}({\bf R}_n) ] | \ ,
\label{eq:gfmc:imp}
\end{equation}
where the sums run over the spin-isospin states $\alpha$, and $\Psi_{T,p}$
is the trial wave function evaluated with a specific choice of pair operator 
orders $p$.
Here $\epsilon$ is a small constant ($\approx 0.01$) that ensures a 
positive-definite importance function so that diffusion can take place 
across nodal surfaces.

The GFMC method allows one to compute ``mixed'' expectation values:
\begin{eqnarray}
\langle O \rangle_{Mixed} & = & \frac{\langle \Psi_{T} | O | 
\Psi(\tau)\rangle}{\langle \Psi_{T} | \Psi(\tau)\rangle} \ , \\
&=& \frac{ \int d {\bf R}_n \Psi_{T}^{\dagger}({\bf R}_{n}) O \Psi({\bf R}_{n},\tau) }
{ \int d {\bf R}_n \Psi_{T}^{\dagger}({\bf R}_{n}) \Psi({\bf R}_{n},\tau) } \ .
\label{eq:omix}
\end{eqnarray}
Because $H$ commutes with the propagator, $\langle{H}(\tau)\rangle_{Mixed}$
is an upper bound to $E_0$ and approaches $E_0$ from above.
However expectation values of operators that do not commute with $H$
are extrapolated using
\begin{eqnarray}
\langle O (\tau)\rangle & = & 
\frac{\langle\Psi(\tau)| O |\Psi(\tau)\rangle}{\langle\Psi(\tau)|\Psi(\tau)\
\rangle} \nonumber \ ,\\
& \approx & \langle O (\tau)\rangle_{Mixed} + [\langle O (\tau)\rangle_{Mixed} 
- \langle O \rangle_T] \ ,
\label{eq:pc_gfmc}
\end{eqnarray}
where
\begin{equation}
\langle O \rangle_T = 
\frac{\langle\Psi_{T}| O |\Psi_{T}\rangle}{\langle\Psi_{T}|\Psi_{T}\rangle} \ .
\end{equation}

In the following we address refinements which have been made to the
GFMC algorithm since Ref.~\cite{PPCPW97}. These refinements are important
in order to make calculations of larger nuclei feasible.

\subsection{Constrained Path Algorithm}

Diffusion or Green's function Monte Carlo
simulations of many-fermion systems generally suffer from the so-called
``fermion-sign problem''.  In essence this results from stochastically
evaluating matrix elements of the form encountered in Eq.~(\ref{eq:omix}).
The Monte Carlo techniques used to calculate the path integrals 
leading to $\Psi({\bf R}_n,\tau)$ involve
only local properties, while antisymmetry is a global property.
This leads to integrands in Eq.~(\ref{eq:omix}) that have oscillating signs 
at large $\tau$, which cause the statistical error to grow exponentially 
with imaginary time.  
The problem is not insurmountable for light systems, because the states
are fairly well separated and one can propagate for a substantial
imaginary time without having unacceptable statistical errors.  
However, the sign problem also grows exponentially
with particle number, as the interchange of any pair of nucleons 
causes a change in sign for the matrix element.  

The sign problem significantly limits the maximum $\tau$ that we can use in 
the simulations;
hence we invoke an approximate technique to deal with it.
The approximation involves keeping only a subset of the paths in evaluating 
the integrals,  and using the knowledge gained in the VMC calculations
to choose the subset of paths.  
It is closely related to methods used previously in 
condensed matter and elsewhere; they generally go by the name of 
constrained-path techniques~\cite{ZCG95,ZCG97}.
Some details of the algorithm, however, are special to the nuclear physics 
case.  

The basic idea of the constrained-path method is to discard those 
configurations that, in future generations, will contribute only noise to 
expectation values.  
If we knew the exact ground state $| \Psi_0 \rangle$, we could
discard any configuration for which:
\begin{equation}
\Psi({\bf P}_n)^\dagger \Psi_0({\bf R}_n) = 0 \ ,
\label{eq:gfmc:const_config}
\end{equation}
where a sum over spin-isospin states is implied.
The sum of these discarded
configurations can be written as a state $| \Psi_d \rangle$,
which obviously has zero overlap with the ground state
\begin{equation}
\langle \Psi_d | \Psi_0 \rangle = 0 \ .
\label{eq:gfmc:const_wvfn}
\end{equation}
The $\Psi_d$ contains only excited states and should decay away as 
$\tau \rightarrow \infty$, thus discarding it is justified. 
However, in  general the ground state $\Psi_0$ is not known exactly
and hence the constraint is imposed approximately using
$\Psi_T$ in the place of $\Psi_0$.

In Green's function or Auxiliary Field Monte Carlo (AFMC)~\cite{CK79,KLV62,BSS81},
the overlap
of the configuration with the trial state evolves smoothly with time. 
The change of the configurations $\Psi({\bf P}_n)$ per time step  
scales with $\sqrt {\Delta \tau}$, which can be made
arbitrarily small.  
If the wave function of the system is a purely real scalar quantity, 
any configuration which yields a negative overlap must first pass 
through a point at which $\Psi_T$ and hence the overlap is zero. 
Discarding configurations at this point is sufficient to stabilize 
the simulation and produce an approximate solution $\Psi_C$ to the 
many-fermion problem.  It solves the many-body Schr\"{o}dinger equation 
with the boundary conditions imposed by the nodes of $\Psi_T$,  
and is known as the fixed-node 
approximation \cite{A76,CA80,MSLK82,RCAL82}.
The approximate $\Psi_C$ is the best wave function
(in the sense of lowest energy) with the same nodes as $\Psi_T$.
The discarded configurations are orthogonal
not only to the trial state but also to the solution
of the constrained problem $\Psi_C$, and it has been shown
that this method produces an upper bound to the ground state
energy~\cite{MSLK82}.

More generally, and particularly in nuclei, 
the trial wave function $\Psi_T$ is a vector in spin-isospin space,
and there are no coordinates for which all the spin-isospin
amplitudes are zero.  
This is also true in AFMC, where the propagated configurations describe 
a fully antisymmetric product of single-particle wave functions.  
In both these cases, the evolution of the wave function remains smooth.
In AFMC it is still possible to use a constraint
of the form given in Eq.~(\ref{eq:gfmc:const_wvfn}).
However, the discarded configurations, while orthogonal to the
trial state $\Psi_T$, are not necessarily orthogonal to the
the solution in the constrained space $\Psi_C$.  
For this reason, the method does not produce an upper
bound to the true ground-state energy~\cite{CGOZ99}.

A new difficulty specific to the nuclear problem is that the overlap 
$\Psi_{T,p}({\bf R}_n)^{\dagger} \Psi({\bf P}_n)$ 
is complex, and is estimated stochastically with a randomly selected 
order, denoted by the subscript $p$, of the operators in $\Psi_T$, Eq.~(\ref{eq:psitgfmc}).  
These overlaps do not evolve smoothly and pass through zero. 
Therefore we can satisfy 
the constraint Eq.~(\ref{eq:gfmc:const_wvfn}) 
only on the sum of discarded configurations ($\Psi_d$),
but not for individual configurations as in Eq.~(\ref{eq:gfmc:const_config}).

The fluctuations in $\Psi_{T,p}$, due to the sampling of the
pair operator orders, are
small compared to the wave function itself, so we define
an algorithm for discarding configurations which resembles
as much as possible the fixed-node or constrained-path
algorithms described above.  
Configurations in the GFMC are obtained with probability
proportional to the importance function $I_{T,p} = I[ \Psi({\bf P}_n), \Psi_{T,p}]$,
Eq.~(\ref{eq:gfmc:imp}), which depends upon $p$.  For each of them the 
overlap $O_{T,p}$ is defined as: 
\begin{equation}
 O_{T,p} = {\rm Re} [ \Psi({\bf P}_n)^\dagger \Psi_{T,p}({\bf R}_n) ] \ ,
\end{equation}
so that the required constraint is  
the sum over discarded configurations of $O_{T,p}/ I_{T,p} = 0$.
We define a probability $P [\Psi({\bf P}_n),\Psi_{T,p}({\bf R}_n)]$
for discarding a configuration in terms of the ratio $O_{T,p}/I_{T,p}$:
\begin{equation}
\begin{array}{rclrrcccl}
P [ \Psi({\bf P}_n), \Psi_{T,p}({\bf R}_n) ]  & = &  0   & &  O/I   &>& \alpha_c \\
   & = &  \frac{\alpha_c -  O/I }{\alpha_c - \beta_c}\hspace*{0.5in}  \alpha_c &>&  O/I   &>& \beta_c \   \\
   & = & 1      \      & &  O/I &<&  \beta_c  \ .
\end{array}
\end{equation}
According to this algorithm configurations with $O/I$ less than $\beta_c$
are always discarded, configurations with $O/I$ greater than $\alpha_c$
are never discarded, and there is a linear interpolation in between.
The values of $\alpha_c$ and $\beta_c$ are held constant during
the simulation. At each step, a configuration is discarded 
with a probability that depends upon the order of pair operators $p$. 
Hence an unbiased estimate of the overlap $\langle \Psi_d | \Psi_T \rangle$
is required.  It is obtained by choosing independent samples $p'$ of the
pair operator order to compute the overlap:
\begin{equation}
\langle \Psi_d | \Psi_T \rangle =  \sum_{\rm \stackrel{\scriptstyle discarded}{\scriptstyle configurations}}
O_{T,p'} / I_{T,p'} \ .
\end{equation}
One or two preliminary runs are sufficient to adjust
the constants $\alpha_c$ and $\beta_c$ such that this overlap is zero
within statistical errors.
When this condition is satisfied,
it is trivial to show that the growth estimate
of the energy, obtained from the growth or decay of
the population with imaginary time, is identical to the 
mixed estimate.  These two energy estimators are
described in detail in Ref.~\cite{PPCPW97}.

Note that in absence of the term with $\epsilon$ in the importance 
function $I_{T,p}$, Eq.~(\ref{eq:gfmc:imp}), the ratio $O_{T,p}/I_{T,p}$ 
is $\pm 1$.  
In constrained path calculations, 
a significantly larger value of $\epsilon$, typically 0.15, is used
to reduce the fluctuations in $O_{T,p}/I_{T,p}$. 
For light nuclei, we have tried a variety
of $\alpha_c$ and $\beta_c$ adjusted to the constraint
without finding any statistically significant differences in the results.
Indeed, even preliminary estimates of $\alpha_c$ and $\beta_c$
which do not precisely satisfy the constraint condition
Eq.~(\ref{eq:gfmc:const_wvfn}) yield results indistinguishable
from the results with finely tuned constraint parameters.
Typical values of $\alpha_c$ and $\beta_c$ for $A$~=~8 are 0.1 and 0.2 
respectively.

In principle, two parameters are not needed to adjust
the constraint to yield zero overlap.  We were motivated to try this
approach by the fact that, in standard fixed-node calculations,
one defines a propagator which goes exactly to zero at the node.
This results in some configurations being discarded which have
small positive overlaps, as well as all configurations which
have negative overlaps.  In the fixed node scheme, this produces
a result of higher order in the time step than simply discarding
configurations with negative overlaps.  For this reason 
both $\alpha_c$ and $\beta_c $ are chosen to be greater than zero. 
The algorithm described above is related to one
proposed by Sorella in the context of condensed matter
simulations~\cite{S98}, but simpler in that it uses only the overlap
and the constraint parameters are held fixed during the simulation.

As discussed in the next subsection this method yields results 
with stable statistical errors independent of $\tau$.  For $A \leq 7$, 
the calculated energy is very close to that obtained without constraints 
in most cases; however in some cases, particularly those with poor 
$\Psi_T$'s, the calculated energy can be significantly below or above 
the true unconstrained result.  The amplitudes of the excited states 
$\Psi_i$, with energies $E_i$, in $\Psi_T$ decay 
exponentially as $\exp(-(E_i - E_0)\tau)$ in the unconstrained $\Psi(\tau)$
without any change in their phase.  
Therefore these give a positive contribution to the mixed energy, making
it ${\geq}E_0$.  The equality is obtained in the limit $\tau{\rightarrow}0$.
The constrained energy can 
have additional errors if, due to a poor choice of 
$\Psi_T$, high-energy excitations get reintroduced in 
$\Psi(\tau)-\Psi_d$.  
Such excitations do not necessarily have the same phase as in $\Psi_T$
and can give contributions to the constrained $E(\tau)$ of either sign.
Such errors can be easily detected by propagating 
$\Psi(\tau)-\Psi_d$ without constraint for a limited number of steps $n_u$. 
Most of the $\Psi_T$ dependence of the calculated energy is eliminated by fairly 
small $n_u$ without a significant increase in the statistical error. 
Thus the wave function $\Psi_C (\tau,n_u)$ used to estimate energies
and other observables in the present calculations, is obtained from:
\begin{equation}
\Psi_C (\tau = n\triangle\tau,n_u) = \{ \exp [ - H n_u\triangle\tau]  \}_u
\{ \exp [ - H (n-n_u)\triangle\tau] \}_c \Psi_T \ ,
\label{eq:gfmc:extrasteps}
\end{equation}
where $\{\cdots\}_c$ signifies propagation with the constraint
and  $\{\cdots\}_u$ indicates normal propagation without constraints.
As is shown in the next subsection,
the simulations are stable for arbitrary $n$, and 
a fairly small $n_u$ ($n_u \triangle\tau \leq 0.01$ MeV$^{-1}$) 
is sufficient to eliminate the $\Psi_T$ dependence.
Evaluation of matrix elements of this wave function can be implemented
very easily on the computer by, after each propagation step, 
labeling those configurations which
are to be discarded and then retaining them in the simulation for
$n_u$ more steps.  In principle
this could be used to evaluate the overlap of the discarded
configurations with $\Psi_T$ after the unconstrained propagation.
This could result in better values of the constraint
parameters, however we find no change in the overlap for the
values of $n_u$ we have considered.

\subsection{Tests of Constrained Path}

We have tested the constrained path algorithm in a variety
of light nuclear systems, studying the dependence upon constraining
wave function and also the convergence of the results obtained
by relaxing the constraint.  
In some cases bad trial wave functions were used to test the
algorithm under extreme conditions.  
In this subsection we present results for
the $^6$Li and $^8$He ground states 
and for 8 neutrons bound in an external well.  
The tests for $^6$Li and 8 neutrons were made using $H=H^\prime$,
Eq.~(\ref{eq:hgfmc}), with no three-body potential.
This eliminates uncertainties from the extrapolation
of $H-H^\prime$, and allows us to have just the exact two-body propagator, $g_{ij}$.  Eliminating the three-body potential also results in faster
calculations allowing smaller statistical errors to be achieved.
In the following all energies are in MeV and imaginary times in MeV$^{-1}$;
for simplicity the units are omitted in most cases.

Figure~\ref{fig:e-of-tau_6Li-test} shows  the ground state energy of $^6$Li 
calculated with
two choices of $\Psi_T$ [Eq.~(\ref{eq:psitgfmc})]:
(1) $\Psi_t$, which is
the full wave function with no $\tilde{U}^{TNI}_{ij;k}$ because
there is no $V_{ijk}$,
and (2) $\Psi_\sigma$, a simplified wave function
obtained by removing the tensor pair correlations leaving only
spin-spin and isospin correlations.  
The subscript $t$ on  $\Psi_t$ emphasizes that it has the essential tensor 
correlations.
In this and the following two figures, constrained-propagation results
with zero and finite $n_u$ are shown as open symbols while solid
symbols show completely unconstrained results or unconstrained continuations
of constrained propagations.  Averages of the last few energies are
shown by solid lines with the corresponding statistical errors indicated
by the surrounding dashed lines; the length of the line indicates the
range of $\tau$ included in the average.
Consider first the results using the full ($\Psi_t$) wave function shown 
in the upper part of Fig.~\ref{fig:e-of-tau_6Li-test}; these
are typical of our production calculations.  
The solid circles show an unconstrained propagation
from $\tau$~=~0 to $\tau$~=~0.1 of 200,000 configurations; the statistical
errors on the last points are growing rapidly, but the energy seems to 
be fairly independent of $\tau$ beyond 0.05.  The average from 
$\tau$~=~0.06 to $\tau$~=~0.1 is --28.16(12), as shown by horizontal 
solid and dashed lines.  It should be close to the correct answer for this 
Hamiltonian. The open triangles show constrained results with $n_u$~=~0 
for the same case. 
This simulation is stable with much smaller statistical errors than
the unconstrained calculation and could
be extended to an arbitrarily large $\tau$; here we find
no significant change after $\tau$~=~0.05. The average over
0.14~$\leq\tau\leq$~0.20 is --27.94(2).
We released the constraint at $\tau$~=~0.20 and continued the propagation
to $\tau$~=~0.27, as shown by solid triangles.  
The errors grow, and the average over this entire region is --28.15(5).
It is likely that this result, which is consistent with the unconstrained
result of --28.16(12) but with smaller statistical error, is the most
accurate available.
A constrained propagation with $n_u$~=~20 
is shown by the open squares; only 50,000 configurations were used.
This simulation is very similar to the $n_u$~=~0 one; the average
is --28.01(5).  
These results suggest that constrained propagation with our typical $\Psi_t$,  
using $n_u$~=~20, may lead to an error of +0.14(7) (or $\sim$0.5\%)
in the binding energy of $^6$Li.

The results in the lower part of Fig.~\ref{fig:e-of-tau_6Li-test} were obtained using
the $\Psi_\sigma$ trial wave function.  This is an extremely bad $\Psi_T$
because, with no tensor correlations, this wave function results in an
identically zero expectation value for the tensor potentials.  Thus
while the $\tau$~=~0.0 expectation value for $\Psi_t$ is --23.95(4),
that for $\Psi_\sigma$ is +31.1(1).  The solid circles show an unconstrained
propagation of 400,000 configurations for this $\Psi_\sigma$ case; the GFMC
has managed to improve the $\tau$~=~0.0 energy by 59 MeV to -28.0(2), but the 
statistical errors are large even after using twice the number of configurations as for the unconstrained propagation with $\Psi_t$.
The open triangles show $n_u$~=~0 constrained propagation of 200,000
configurations.  This propagation becomes stable after $\tau$~=~0.12,
but it has converged to the significantly overbound result of --30.2(1).
The solid triangles show the results of releasing the constraint at
$\tau$~=~0.20; the energy immediately goes up to the correct value
with an average of --28.0(2).
Finally the open squares show $n_u$~=~20 constrained propagation of 400,000
configurations.  The repeated use of 20 unconstrained steps gives
a stable propagation that is quite accurate; the average is --28.27(7) 
with an error of --0.12(9).
The unconstrained continuation of this calculation, shown by the
solid squares, makes no significant change.
These results show the need to have unconstrained steps before 
evaluation of the energy or any other observable.

The second test case is a system of 8 neutrons bound
in an external one-body potential. We have previously reported results
for such systems as a basis for comparing Skyrme models with microscopic
calculations based on realistic interactions~\cite{PSCPPR96}.
As in the previous example, the neutrons interact via the $v^{\prime}_{ij}$, with no
three-nucleon interaction.
Because systems of neutrons are not self binding, the Hamiltonian
also includes an
external one-body potential of Woods-Saxon form,
\begin{equation}
V_1 (r) = \sum_i \frac{V_0}{1 + \exp [ - (r_i - r_0)/a_0 ]} \ ;
\end{equation}
the parameters are $V_0 = -20$ MeV, $r_0 = 3.0$ fm, and $a_0 = 0.65$ fm.
Neither the external well nor the internal $v^{\prime}_{ij}$ potential are
individually attractive enough to produce a bound state of eight
neutrons, however the combination does produce binding.

The Jastrow $\Psi_J$ for neutron drops is just
\begin{equation}
     |\Psi_J\rangle = \prod_{i<j}f_c(r_{ij}) |\Phi_A(JM)\rangle \ .
\end{equation}
where $\Phi_A$ is a Slater determinant of single-particle orbitals.
We compare results obtained with the full $\Psi_T$, 
again refered to as $\Psi_t$ because of the tensor operators,
shown in the upper part of Fig.~\ref{fig:e-of-tau_8n-test}, 
with those obtained with the $\Psi_J$, shown in the lower part.  

The solid circles show results of unconstrained
propagation using 400,000 configurations of the $\Psi_t$ 
and 560,000 configurations of the $\Psi_J$ trial wave functions, respectively.
The errors are growing rapidly by $\tau$~=~0.06, and the calculations
do not appear to have converged.  Averages of the last four energies are 
--39.3(1) and --38.9(1) for $\Psi_t$ and $\Psi_J$ respectively, 
but these are clearly just upper bounds.
Constrained propagations using the $\Psi_J$ trial wave function and $n_u$~=~0 and
$n_u$~=~20 are shown by the open triangles and squares, respectively
in the lower part of the figure.  These
have stabilized beyond $\tau$~=~0.1, with end averages of 
respectively --39.58(4) and --39.79(6).  
Both propagations were continued without constraint; their results are 
shown by solid triangles and squares.
The energy remains stable but the errors grow rapidly; the averages are
--39.86(11) and --39.85(34).  
In this case constrained propagation with $n_u$~=~20 gives results
that are below the best unconstrained 
(starting from ${\tau}=0$)
upper bound with $\Psi_t$ by 0.5(1).
The constrained result is consistent with unconstrained continuations
beyond ${\tau}=0.25$, indicating its reliability.

A different situation obtains with the variationally better $\Psi_t$ trial wave function;
the $\tau$~=~0, VMC energy for $\Psi_t$, --35.30(4), is lower than
the --30.62(9) given by $\Psi_J$, but the difference is not as large as in 
the case of $^6$Li.  
Constrained propagation using $\Psi_t$ and $n_u$~=~0 is shown by the 
open triangles in the upper part; it stabilizes beyond $\tau$~=~0.1, but at a value that is
1 MeV too high.  
The results of releasing the constraint are shown by the solid triangles;
the energy immediately drops 1 MeV and the final average is --39.71(10).
Finally constrained propagation for $\Psi_t$ and $n_u$~=~20 is shown by the 
open squares.  The results are stable with an average of --39.74(5),
in agreement with the best results obtained in the previous paragraph.
Releasing the constraint does not result in any significant change, as is
shown by the solid squares.
This case also confirms the need to release the constraint before 
measurements, with the surprising result that a variationally better 
wave function may not necessarily provide better constraints to guide the GFMC. 
Nevertheless, constrained propagation with $n_u$~=~20 gives results with 
presumably less than 1\% error, for both trial wave functions.

Finally, in Fig.~\ref{fig:e-of-tau_8He-test} 
we show results for the $^8$He ground state
with the full Argonne $v_{18}$ plus Urbana IX three-nucleon interaction.
We are presenting this case because we find that $^8$He is the most
difficult nucleus we have studied, 
in terms of obtaining reliable error estimates,
and the need for $n_u > 0$.
All of the calculations are made with the full 
$\Psi_T$ of Eq.~(\ref{eq:psitgfmc}).
In this figure the results of a standard calculation without
constraint (solid circles) are compared to those with constrained-path 
$n_u$~=~0 (open triangles) and $n_u$~=~20  (open squares), 
and their unconstrained continuation beyond $\tau$~=~0.2
shown by solid triangles and squares. 
The average unconstrained energy in the $0.03 \leq \tau \leq 0.06$ 
range, --26.1(3), is clearly an unconverged upper bound.  
The $n_u$~=~0 and 20 end averages, --26.89(9) and --27.16(15), 
are below it by $\sim$1 MeV. 
The averages of the results obtained after the constraint is released 
at $\tau$~=~0.2, --27.5(4) and --26.9(2), are not significantly different
from the constrained averages.

From these, and many other, tests we conclude that constrained 
propagation, including unconstrained steps prior to measurement, yields
results that are reliable.  In the $A=8$ nuclei and neutron drops 
they are up to 4\% below the unconverged upper bounds 
that can be obtained by unconstrained propagation up to $\tau = 0.06$, and have 
smaller statistical errors.  The constrained path results with 
$n_u \sim 10$ to 20 are stable within 1\% with respect to reasonable, 
and in case of $^6$Li even unreasonable, changes in $\Psi_T$ used to 
constrain the paths.  Our present practice is to use the full $\Psi_T$, 
Eq.~(\ref{eq:psitgfmc}) in constrained path calculations with $n_u$~=~20 for the 
neutron-rich He isotopes and neutron drops and $n_u$~=~10 for all other nuclei
where even $n_u = 0$ calculations seem to be fairly accurate.

\subsection{Resonance states in constrained-path algorithms}

Since the constrained-path algorithm 
is stable to large imaginary time, it is useful to
consider the large-$\tau$ behavior of the energy of 
resonant states.   The bound-state simulations
described above yield asymptotically stable energies out to very large
imaginary times.  Resonant states,
however, are more delicate.  In principle these states
should decay to separated clusters, and in fact this does
occur with the constrained-path algorithm.  The rate of
this decay presumably depends not only on the resonance energy 
but also on the width of the state, and the trial state wave function
used to impose the constraint.

We have studied several cases including
the two unbound p-wave states  in $^5$He. 
The GFMC energy consistently decreases with $\tau$ for both
the $J^\pi = 1/2^-$ and the $3/2 ^-$ states.  The 1/2$^-$
state, lying higher in energy and with a larger width, decays
more quickly than the $3/2^-$ state.  We have verified that
the system breaks apart into a separated $\alpha$ plus $n$
by studying the $pp$ pair distribution function, which 
depends only upon the internal structure of the $\alpha$-particle.
This distribution remains constant out to the largest imaginary
times (3 MeV$^{-1}$) studied, 
while the $pn$ and $nn$ distributions steadily become broader.

We have also studied three low-lying states in $^6$Li out to
large $\tau$.  The $J^\pi = 1^+$ ground state is stable,
the $3^+$ excitation is a narrow resonance, while the
$2^+$ state lies higher in energy and is broader.  The GFMC energies
of these states are displayed in Fig.~\ref{fig:e-of-tau_6Li-largetau} 
as a function of $\tau$.
The ground state is clearly stable out to the largest imaginary
times.  The narrow $3^+$ shows a plateau in $E(\tau)$, decreasing
only modestly in energy, though of course it will eventually also
decay to the $\alpha$+d threshold energy.  
The $2^+$ state is much broader and decays much more
quickly in imaginary time.  Therefore, comparison of
the energies obtained for such broad states with experiment
could be misleading.
The $\alpha$+d threshold energy for this Hamiltonian is shown in 
the figure; it is clear that both excited states are far from
convergence to this energy.

  It is possible to compute energies that may be directly compared 
with experiment for resonant states
that have only a single two-body channel available for
breakup~\cite{CPW84}.  Scattering observables, including scattering length, 
effective range, and phase shifts, can be calculated
directly by imposing scattering boundary conditions on the
asymptotic wave function. This is quite important for studying
a variety of interesting low energy reactions,
including parity-violation and important astrophysical
reactions, but has not been considered in this work.

\subsection{Three-body Propagator}

The small-time-step propagator of Eq.~(\ref{eq:fullprop}) involves two complete
evaluations of the three-body potential (two sums over all triples) and one
product of all pair propagators for each time step.  Thus as the number of
nucleons is increased, the time spent in the three-body part of the propagation
becomes a larger and larger fraction of the total time for the calculation.
For this reason it is desirable to find a less costly treatment of the
three-body propagator.

It was noted in Eq.~(\ref{eq:psitgfmc}) that the full $U^{TNI}_{ijk}$
can be replaced with a $\tilde{U}^{TNI}_{ij;k}$ that omits the commutator
part of $V^{2\pi}_{ijk}$ with very little degradation of the variational energy,
and that the resulting correlation is an operator in the spins and
isospins of only two nucleons.  This led us to consider the combined two- and
three-body propagator (we omit the spin-isospin indices):
\begin{eqnarray}
\tilde{G}({\bf R},{\bf R}^{\prime})& = &e^{E_{o}\triangle\tau}
{\cal S}\prod_{i<j}\frac{\tilde{g}_{ij}({\bf r}_{ij},{\bf r}_{ij}^{\prime})}
{g_{0,ij}({\bf r}_{ij},{\bf r}_{ij}^{\prime})} \\
\tilde{g}_{ij}({\bf r}_{ij},{\bf r}_{ij}^{\prime}) &=& 
\left [1 - \frac{\triangle\tau}{2} \sum_{k \neq i,j} 
\alpha V^{2\pi,A}_{ij;k}({\bf R}) \right ]
~ g_{ij}({\bf r}_{ij},{\bf r}_{ij}^{\prime}) ~
\left [1 - \frac{\triangle\tau}{2} \sum_{k \neq i,j} \alpha V^{2\pi,A}_{ij;k}({\bf R}^{\prime})\right ]
                                                                 \nonumber \\
& & \times \exp \left[{- \frac{\triangle\tau}{2} 
\sum_{k \neq i,j} (V^{R}_{ijk}({\bf R})+ V^{R}_{ijk}({\bf R^{\prime}}))} \right] \ .
\end{eqnarray}
The $V^{2\pi,A}_{ij;k}$ is defined in Eq.~(\ref{eq:2piA}) as the anticommutator part 
of $V^{2\pi}_{ijk}$, and $\alpha$ is chosen so that 
\begin{eqnarray}
\alpha \langle V^{2\pi,A} \rangle = \langle V^{2\pi} \rangle \ ;
\end{eqnarray}
typically $\alpha = 1.6$.
This $\tilde{g}_{ij}$ can be reduced to a single $4 \times 4$ operator
in $i,j$ spin space for each of the two isospin states, and thus
takes little more time to evaluate than just $g_{ij}$.
Propagation with just $\tilde{G}$ gives much the same results as
with the much more costly propagator of Eq.~(\ref{eq:fullprop}), but
we attempt to make a more reliable propagation by using the following sequence
as the basic propagation step:
\begin{eqnarray}
\tilde{G}({\bf R}_{i+n},{\bf R}_{i+n-1}) \tilde{G}({\bf R}_{i+n-1},{\bf R}_{i+n-2}) \cdots
\tilde{G}({\bf R}_{i+n/2+1},{\bf R}_{i+n/2}) \nonumber \\
\times \{1 - n\triangle\tau \sum_{i<j<k}[ V^{2\pi}_{ijk}({\bf R}_{i+n/2}) - 
\alpha V^{2\pi,A}_{ij;k}({\bf R}_{i+n/2}) ]\} \nonumber \\
\times \tilde{G}({\bf R}_{i+n/2},{\bf R}_{i+n/2-1}) \cdots \tilde{G}({\bf R}_{i+1},{\bf R}_i) \ ,
\label{eq:gcorrection}
\end{eqnarray}
where $n$ is a small (typically 4) number of steps.  
Here we go from ${\bf R}_i$ to ${\bf R}_{i+n}$ by making $n/2$ steps
with $\tilde{G}$; then applying $n$ times the correction due to the
difference of the complete $V^{2\pi}$ and the approximate 
$\alpha V^{2\pi,A}$, both computed at the position ${\bf R}_{i+n/2}$;
and finally making another $n/2$ steps to ${\bf R}_{i+n}$.

Table~\ref{table:newprop} shows the reliability of this method for $^6$Li(gs).
The first line gives results with the old method, Eq.~(\ref{eq:fullprop}), and
the next three lines use the new method with increasing values of $n$ [$n=\infty$
means that no correction for the approximation in $\tilde{G}$ 
is made with Eq.~(\ref{eq:gcorrection})].
The last line shows results using just a two-body propagator; that is, the three-body
potential is included only perturbatively.  Clearly a three-body propagator
is essential for obtaining results with $< 1\%$ error, 
however the $\tilde{G}$ is reliable to better than 1\%.

\subsection{Numerical Evaluations}

The p-shell GFMC
calculations reported here were made with constrained-path propagation
to $\tau = 0.2$~MeV$^{-1}$ using 400 steps of $\triangle\tau =
0.0005$~MeV$^{-1}$.  Energies and other observables were evaluated
every 20 steps and the last 7 ($\tau \geq 0.14$~MeV$^{-1}$) values were
averaged.  At least 10 unconstrained steps were taken before the
observables were computed (20 steps were used for $^8$He).  The $A = 8$
results used 10,000 to 20,000 initial configurations.  Once the
propagation has stabilized (typically by $\tau = 0.1$~MeV$^{-1}$), the
constraint removes from 0.7\% [for $^8$Be(gs)] to 1.7\% [for
$^8$Be(3$^+$)] of the configurations at each propagation step (the
percentages for $^8$He and $^8$Li are in this range).  The removed
configurations are replaced by new ones generated by branching with the
average branching probability chosen to maintain an approximately
constant population.

Most of the GFMC calculations reported here were made on the 128-node
SGI Origin 2000 in the Mathematics and Computer Science division of
Argonne National Laboratory.  The individual nodes in this machine are
250 MHz R10000 processors which (when the processors are not being
shared with other users and memory remains local to each processor)
deliver sustained speeds of $\geq 200$ MFLOPS for the 8-nucleon
calculations.  This results in a 10,000 configuration calculation for
$^8$Be taking about 460 node-hours.  The approximate times required for
other nuclei can be determined from Table \ref{table:scaling}.  The
columns of this table show the number of nucleons, the number of pairs,
and the size of the spin-isospin vector.  We find that the total
calculational time is proportional to the product of these three
numbers; this product (scaled to 1 for $^8$Be) is given in the last
column.

Typically we have computed the $M = J$ state of a given nucleus; this
allows us to directly evaluate the spectroscopic quadrupole moment and
magnetic moment.
Recently we have realized that if we compute the $M = 0$ state of even-$A$
nuclei, the size of the spin-isospin vector can be reduced by a factor
of two.  This is done by observing that for even $A$,
\begin{equation}
  \Psi_{-s_1,-s_2,\cdots,-s_A}(J,M=0) = (-1)^{\case{1}{2}A + J - M_S}
    \Psi^*_{s_1,s_2,\cdots,s_A}(J,M=0) \ ,
\end{equation}
where $s_i$ is the spin-projection of nucleon $i$ and $M_S = \sum s_i$.
Thus only half of the spin-isospin vector needs to be computed and stored,
resulting in a saving of half the computer time for even $A \geq 8$ nuclei. 
This saving is not included in the times discussed in the previous paragraph.
Unfortunately, the corresponding relation in odd-$A$ nuclei relates
$\Psi_{-s_i}(J,M)$ to $\Psi_{s_i}(J,-M)$ and thus does not reduce
the computational effort.

\section{ENERGY RESULTS}

\subsection{Ground-State and Excitation Energies}

In this Section we present VMC and GFMC results for the Hamiltonian
consisting of Argonne $v_{18}$ plus Urbana IX.
Figure~\ref{fig:e-of-tau_8Be} shows the $E(\tau)$ for the lowest
$T = 0$, $J^\pi = 0^+, 2^+, 4^+, 1^+, 3^+$ states of $^8$Be.  The
solid and dashed lines show the average energy and its statistical
error; these numbers are reported in Tables \ref{table:energy} and 
\ref{table:excited}.  
The $E(\tau)$ are
\begin{equation}
E(\tau) = \langle H^\prime(\tau) \rangle_{Mixed}
+ \langle H(\tau) - H^\prime(\tau) \rangle \ ,
\end{equation}
where the $\langle H - H^\prime \rangle$ is perturbatively extrapolated
by Eq.~(\ref{eq:pc_gfmc}).
Figures \ref{fig:e-of-tau_8Li} and 
\ref{fig:e-of-tau_8He} show the corresponding $^8$Li and $^8$He results.
In all cases for which the states are experimentally narrow or stable,
the $E(\tau)$ rapidly decreases for small $\tau$ and stabilizes before
$\tau = 0.1$ MeV$^{-1}$.  It is clear that $E(\tau)$ has not converged
by $\tau = 0.2$ MeV$^{-1}$ for the experimentally very broad
$^8$Be(4$^+$) state, making it not possible to determine an accurate
excitation energy for this state.  
The other broad states [$^8$Be(2$^+$)
and $^8$He(2$^+$)], and the experimentally unknown $^8$He(1$^+$) state
seem reasonably converged.

Table~\ref{table:energy} shows the computed and experimental
ground-state energies.  The errors shown in parentheses are only the
Monte Carlo statistical errors; the systematic errors discussed in the
previous section and in Ref.~\cite{PPCPW97} could add an additional 1\%
to the GFMC error.  This paper is our first formal publication of $A$~=~8
results.  Results for $A = 6$ and 7 were published in
Ref.~\cite{PPCPW97}; most of the corresponding values in Table
\ref{table:energy} have been recomputed using improved VMC wave functions
and GFMC propagation.  Most of the VMC values have not changed
significantly from those in Ref.~\cite{PPCPW97}; the exception is $^7$He
which is now constrained to what we think is a more reasonable rms
radius and actually has a higher energy.  The $A$~=~6 and 7 GFMC
energies have now all been computed with constrained propagation to
$\tau = 0.2$~MeV$^{-1}$ instead of the unconstrained propagation to
only $\tau = 0.06$~MeV$^{-1}$ used in Ref.~\cite{PPCPW97}.  This has
resulted in a lowering of $^{6,7}$He energies by 0.47(17) and 0.63(23) 
MeV respectively; other energies changed by less than 1\%.  
For $A$~=~8 nuclei, the GFMC improves the $\Psi_T$ energy by $\sim$10~MeV;
our best $\Psi_V$ provides only $\sim$1.5~MeV of this improvement.

Table \ref{table:excited} shows excitation energies for the $A$~=~8
nuclei.  The errors are the combined statistical errors of the
energies computed for the given state and the ground state.
We see that, while the absolute VMC energies have substantial errors
as compared to the GFMC values,
the VMC excitation energies are typically within an MeV of the GFMC 
values.  

Figure \ref{fig:energies} shows the energies of nuclear states for $4
\leq A \leq 8$ and Fig.~\ref{fig:excitations} shows the corresponding
excitation spectra.  As is the case for the ground-state values, most of
the $A$~=~6 and 7 excited-state energies have been recomputed with
improved VMC wave functions and all of the GFMC propagations have been
made to $\tau = 0.2$~MeV$^{-1}$.  The VMC  excitation energies of the
lowest $^7$Li($\case{1}{2}^-$) and $^7$Li($\case{5}{2}^-$) states
have been significantly reduced from the
values reported in Ref.~\cite{PPCPW97}.  
These are the result of improved mixings of the
different symmetry states in the variational wave function.

The figures confirm the principal $A$~=~6 and 7 results of Ref.~\cite{PPCPW97}
for $A$~=~8: the Hamiltonian consisting of Argonne $v_{18}$ plus Urbana IX underbinds
nuclei in the p-shell with the underbinding becoming worse as one
increases $A$ or $N-Z$.  However the predictions of the excitation
spectra are generally reasonable; the rms error of all of the 16 excitation
energies computed by GFMC is only 650 keV; for the 9 states with experimental
width less than 200 keV it is 540 keV.  Considering that no parameters
were adjusted to fit these p-shell energies, this is quite respectable.
We do note that, where there is an experimental value to compare to,
the VMC excitation energies of second excited states of a given $J^\pi$, which are
very difficult to compute by GFMC, are usually too high.

Although the Argonne $v_{18}$ plus Urbana IX Hamiltonian consistently underbinds
the p-shell nuclei, the errors are small compared to the magnitudes
of the potential energies.  Table \ref{table:gfmc_terms} shows the
perturbatively extrapolated GFMC expectation values for the kinetic energy
and potential energy terms.  
As is discussed in Ref.~\cite{PPCPW97}, the perturbatively extrapolated GFMC 
terms do not add up to the total energy, which is the most reliable number
computed by this method.
The total two-body potential, $v_{ij}$ is 
dominated by the one-pion exchange term, $v^{\pi}_{ij}$.
There is a large cancellation between the kinetic and two-body potential 
energies so that their sum is only 15\% to 20\% of the two-body potential.
The total three-body potential is typically 4\% to 5\% of the two-body potential.  
The repulsive $V^{R}_{ijk}$ typically cancels 45\% of the $V^{2\pi}_{ijk}$.
As an indication of the smallness of the errors produced by this Hamiltonian,
consider the ground state of $^8$Be:  the difference of the computed and
experimental energies is 2.1 MeV; the two-body potential is $-$301~MeV,
$V^{2\pi}_{ijk}$~=~$-$27~MeV, and $V^{R}_{ijk}$~=~+12.3~MeV.  Thus the corrections
that have to be made to $V_{ijk}$ are significantly smaller than the
terms already included.

\subsection{Isobaric Analog States}

Energy differences of isobaric analog states are sensitive probes of the
charge-independence-breaking parts of the Hamiltonian.
To study these it is useful to express the energies in an isobaric multiplet,
characterized by $A$ and $T$, in terms of the isospin multipole operators
of order $n$:
\begin{equation}
   E_{A,T}(T_z) = \sum_{n\leq 2T} a^{(n)}_{A,T} Q_n(T,T_z) \ .
\end{equation}
The $Q_n(T,T_z)$ are orthogonal functions for projecting out isovector,
isotensor, and higher-order terms~\cite{P60}; the first terms are
$Q_0=1$, $Q_1=T_z$, and $Q_2=\case{1}{2}[3T_z^2-T(T+1)]$.
The coefficients $a^{(n)}$ are then obtained from the calculated energies: 
\begin{equation}
   a^{(n)}_{A,T} = \sum_{T_z} Q_n(T,T_z) E_{A,T}(T_z)
                 / \sum_{T_z} Q^2_n(T,T_z) \ ,
\end{equation}
or perturbatively from expectation values of the isomultipole
operators present in the Hamiltonian.

In first-order perturbation theory, the electromagnetic interaction
contributes to the $a^{(n)}$ for $n=0$, 1, and 2, 
the kinetic energy to $n=0$ and 1,
the nuclear CSB potential to $n=1$,
and the nuclear CD potential to $n=2$.
Because a significant portion of the $v^{CD}_{ij}$ comes from one-pion
exchange, there should also be a CD component to $V^{2\pi}_{ijk}$; 
however, a plausible extension of the Urbana IX model gives negligible
contributions of 3 keV or less to the $n=2$ terms.
The $a^{(n)}$ for higher $n$ are zero in first order with our Hamiltonian, and
there is little experimental evidence for $n\geq3$ terms in nuclei~\cite{BK79}.
We have made VMC calculations of the $a^{(1)}$ and $a^{(2)}$ in first order by using
a CI wave function of good isospin,~$T$, and simply varying $T_z$
to compute the $E_{A,T}(T_z)$; in the GFMC calculations we evaluate expectations
of the isomultiplet operators for the $T_{z}=-T$ nuclei.
The CSB and CD parts of the Hamiltonian can induce corresponding changes in the nuclear wave
functions, leading to higher-order perturbative corrections to the splitting
of isospin mass multiplets.
However, it is difficult for us to estimate these higher order effects
reliably in either the VMC or GFMC calculations.

Table~\ref{table:analog} shows results for the $T=1$ and 2 isovector
and isotensor coefficients in $A=8$ nuclei compared to experiment.
In obtaining the experimental $a^{(2)}_{8,1}$ coefficient, we have used the 
average of the two isospin-mixed (2$^+$;0+1) states; our calculation of
this mixing is discussed in the next subsection.
The contributions of the complete $v^{\gamma}$, $v^{CD}$, $v^{CSB}$, 
and $K^{CSB}$ terms to the $a^{(n)}$ are also given.
The results show that the present Hamiltonian and VMC wave function match 
the experimental CSB and CD of the $T=1$ multiplet fairly well, but the 
calculated value for $a^{(1)}_{8,2}$ is about 10\% too small.
The GFMC consistently reduces the coefficients, mostly by reducing
the Coulomb term.  For the isovector  coefficients, this worsens
the comparison with experiment, however the case of $a^{(2)}_{8,2}$ is
significantly improved.
In all these calculations, the Coulomb interaction between protons is the 
dominant contribution to the $a^{(n)}$, but the strong interaction CSB
terms serve to bring the final results closer to experiment.
However, in view of the fact that the $A=8$ nuclei are underbound with the
present Hamiltonian, it is premature to use these calculations as a
precise test of the charge-independence-breaking components of the interaction.

\subsection{Isospin Mixed States}

The (2$^+$;1) isobaric analog of the $^8$Li ground state in $^8$Be is very
close in energy to the second (2$^+$;0) excitation, as shown in 
Fig.~\ref{fig:expt8}, and is in fact experimentally observed to be 
isospin-mixed with it.
There are also fairly close (1$^+$;0,1) and (3$^+$;0,1) pairs at slightly
higher energies in the $^8$Be spectrum.
Our VMC calculations do not get the (2$^+$;0,1) states so close together, 
although the other pairs do come out quite near each other.
However, we can calculate the isospin-mixing matrix elements that connect 
these pairs of states:
\begin{equation}
   E_{01}(J) = \langle \Psi(J^+,0) | H | \Psi(J^+,1) \rangle \ .
\end{equation}
This is done by increasing the model space to include both $T=0$ and 1
components, which corresponds to a wave function vector with 10,752 terms,
and employing the same off-diagonal evaluations used to determine
the $\beta_{LS[n]}$ wave function components.

Results for the $E_{01}(J)$ are given in Table~\ref{table:mixing}.
The experimental values are determined from the observed decay widths
and energies~\cite{B66}.
The dominant contribution, from the Coulomb potential, typically 
accounts for less than half of the matrix element.
We find the magnetic moment part of the electromagnetic interaction and 
the strong CSB interaction can provide a significant boost, although we
still underpredict the experimental mixing by $\sim$20\%.
The spatial symmetry components of the different wave functions, 
as given in the Tables~\ref{table:beta81} and \ref{table:beta80},
for the (2$^+$;0,1) and (3$^+$;0,1) pairs are indeed fairly close to 
each other, since they have similar sizes and signs for the largest components.
However, the (1$^+$;0,1) states are not so similar, particularly due to the
large $T=1$ $^1$P[31] component that is not available to the $T=0$ state.
This may be why the $E_{01}(1)$ is noticeably smaller, which shows up through
the change in sign of the magnetic moment contribution.

\section{MOMENTS AND DENSITY DISTRIBUTIONS}

The proton rms radii, magnetic moments, and quadrupole moments for the $A=8$
nuclear ground states are given in Table~\ref{table:radii}.
The calculations have been made in impulse approximation for both the initial 
VMC wave function and perturbatively after GFMC constrained path propagation.
Comparing the VMC and GFMC radii, it appears that the variational wave 
functions are a little too compact for this Hamiltonian.
However, given that the present interaction underbinds these nuclei, the 
actual radii should be smaller than the GFMC values.
Attempts have been made to determine the matter radius of $^8$He by 
proton-scattering experiments in inverse kinematics.
Interpretation of the data is not model-independent, however, and has resulted
in estimates ranging from 2.45 fm \cite{Alk97} to 2.6 fm \cite{AT98}.
The corresponding VMC and GFMC matter radii from Table~\ref{table:radii} are
2.69 fm and 2.92 fm, respectively.
In the not too distant future, it may be possible to trap $^8$He, $^8$Li, 
and $^8$B long enough to determine their charge radii to high precision
by atomic means.

The magnetic moments of $^8$Li and $^8$B have been measured by 
$\beta$-radiation detection of implanted polarized ions~\cite{R89}.
Our IA calculations should be supplemented by meson-exchange currents (MEC);
their contributions are dominantly isovector, and have been shown to change
the magnetic moments of $^3$H and $^3$He by $\pm$ 0.4 $\mu_N$ \cite{SPR89}.
The isoscalar average of the calculated magnetic moments is close the 
experimental value, and 
it is plausible that MEC contributions will bring both magnetic 
moments into good agreement with experiment.
The quadrupole moments of $^8$Li and $^8$B have been measured by $\beta$-NMR
and nuclear quadrupole resonance techniques~\cite{S97}.
The VMC and GFMC results are fairly close for $^8$Li, and just a little 
above the experimental value.
The calculations for $^8$B are significantly further apart, despite the 
proton rms radii differing by only a small amount; nevertheless they
bracket the experimental value.
We have also calculated the hexadecapole moment for these two ground states,
and find it to be consistent with 0.

One- and two-nucleon density distributions have been calculated in both
VMC and GFMC. 
The GFMC densities are a little more spread
out and less peaked than the corresponding VMC densities, 
as reflected in the charge radii differences;
here we show only the GFMC densities.
Single-nucleon density distributions for the $A=8$ nuclei are shown in 
Fig.~\ref{fig:8bodyrho1}; they are normalized such that the integrated value
equals the appropriate total value of $N$ or $Z$.
The two protons in $^8$He are the most peaked distribution, which should be 
expected on the grounds that they are mostly confined to the $\alpha$ core,
while the six neutrons in $^8$He have the broadest distribution.
The proton distribution in $^8$Li is also fairly peaked near the origin,
but is broader than in $^8$He since there is one additional proton in the
p-shell.
The neutron distribution in $^8$Li is comparable to the $^8$He neutrons
near the origin, is slightly larger in the 1-2 fm range, and then falls
below at larger distances; the intermediate-range excess may be due to
the significantly greater binding of $^8$Li compared to $^8$He.
In contrast, the $^8$Be proton and neutron distributions are much less peaked
at the origin and are rather flat out past 1 fm.
This could be because $^8$Be has a significant 2$\alpha$ component
in its intrinsic structure, with the two $\alpha$'s sitting side-by-side;
as discussed in the next section.

A logarithmic plot of the single-nucleon densities for $^{4,6,8}$He isotopes
is shown in Fig.~\ref{fig:herho1}.
The neutron halos in $^6$He and $^8$He are clearly evident,
while the proton cores of these nuclei are nearly identical.
The peak neutron and proton distributions in $^6$He and $^8$He are much
reduced compared to $^4$He, because of the motion of the $\alpha$ core against
the center-of-mass.
Two-proton density distributions for $^{4,6,8}$He are shown in 
Fig.~\ref{fig:herho2}.
The relative proton-proton density in $^{6,8}$He is not effected by the
motion of the $\alpha$ relative to the center of mass; thus $\rho_{pp}$
is an indicator of changes in the internal structure of the $\alpha$ core
in $^{6,8}$He.
Here we note that there is only a small change in the $\alpha$ core
of $^{6,8}$He, i.e., about a 10\% reduction in the peak compared to $^4$He,
which must be compensated in the long-range tail, since the total integral 
under the curves is unity in each case.
This reduction could be due to swelling of the $\alpha$ core, or due to
charge-exchange interactions between the protons and p-shell neutrons.

We have also calculated the single-nucleon momentum distributions shown in 
Fig.~\ref{fig:8bodyrhok}, using the VMC wave functions and the method of
Ref.~\cite{SPW86}, with some algorithmic improvements.
The $N(k)$ are normalized to $N/(2 \pi)^3$ or $Z/(2 \pi)^3$.
All the distributions show a remarkably similar structure: a low-momentum
core attributable to s- and p-shell orbitals, followed by a high-momentum
tail beyond 2 fm$^{-1}$ that is the sum of many small-amplitude 
higher-orbital contributions.
This high-momentum tail is already evident in the $^4$He momentum 
distribution and is intimately related to the D-state of $^4$He, present
due to the strong tensor forces and corresponding correlations in the
wave function.
The $^8$He proton distribution is very similar to that of $^4$He, showing
that the $\alpha$ core is not much altered, while the neutrons exhibit a p-shell peak 
near 0.5 fm$^{-1}$.
The $^8$Li proton distribution is a little broader, as might be expected
by the addition of a p-shell proton, while the neutrons show a smaller p-shell
peak.
The $^8$Be momentum distribution is almost exactly double that of $^4$He
except at very low momenta, indicating a significant 2$\alpha$
structure.

\section{INTRINSIC SHAPES}

The ground state and first two excited states of $^8$Be have an approximate
rotational energy spectrum, and are assumed to be well approximated as
two $\alpha$'s rotating around their common center of mass.  
This structure is not manifest in the shell model part of the 
VMC wave functions, in which the $^8$Be states are constructed 
from an $\alpha$-like core surrounded by four 
p-shell nucleons coupled to the appropriate total angular momentum.
In this section we describe an attempt to study the intrinsic structure
of the $^8$Be states described by the VMC wave functions including 
correlations.

The standard Monte Carlo method for computing one-body densities, 
$\rho({\bf r})$, is to make a random walk that samples 
$|\Psi({\bf r}_1,{\bf r}_2,\cdots,{\bf r}_A)|^2$ and 
to bin ${\bf r}_1,{\bf r}_2,\cdots,{\bf r}_A$ 
for each configuration
in the walk.  The density is then proportional to the number
of samples in each bin.  In the case of a $J$~=~0 nucleus, this
``laboratory'' density will necessarily be spherically symmetric.

We can attempt to find the intrinsic density in body-fixed coordinates 
by computing the moment of inertia matrix:
\begin{equation}
{\cal M} = \sum_{i=1}^A  \left( \begin{array}{ccc}
  x_i^2   &  x_i y_i   & x_i z_i   \\
 y_i x_i  &   y_i^2    & y_i z_i   \\
 z_i x_i  &  z_i y_i   &   z_i^2   \\
\end{array} \right)    ~~,
\end{equation}
for each configuration.
We then find the eigenvalues and eigenvectors of ${\cal M}$,
rotate to those principal axes, and bin the resulting 
${\bf r}_1^\prime,{\bf r}_2^\prime,\cdots,{\bf r}_A^\prime$.
The eigenvector with the largest eigenvalue is chosen as the
${\bf z}^\prime$ axis; further choices of the $\pm$ direction along 
${\bf z}^\prime$ and which eigenvectors to assign to ${\bf y}^\prime$
and ${\bf x}^\prime$ may also be made or averaged over.
This procedure will not produce a spherically symmetric distribution,
even if there is no underlying deformed structure, because 
almost every random configuration will have  principal axes of different
lengths and the rotation will always orient the longest principal axis
in the ${\bf z}^\prime$ direction.  We have made a number of tests
using simple wave functions with no internal correlations and
also nuclei like $^4$He which, in our models, should have no
intrinsic deformations.  We find that the projected ``intrinsic''
density for such cases is always prolate but that no other
artificial structure is introduced.

When the above procedure is applied to the $^8$Be rotational states,
a dramatic intrinsic structure is revealed.  Figure~\ref{fig:be0-con}
shows this calculation for the ground state.  The figure shows
contours of constant density plotted in cylindrical coordinates,
with the z-axis being the axis of quantization.  The calculation
was made for the VMC wave function.  The left side
of the figure shows the standard density calculation; we can think
of this as the density in the ``laboratory'' frame.  For the $J$~=~0
ground state, this is spherically symmetric as shown.  The right
side of the figure shows the intrinsic density computed as described
above.  In this case the orientation (up or down) along ${\bf z}^\prime$ 
and around ${\bf z}^\prime$ ($r = \sqrt{x^2+y^2}$) was averaged over.
It is clear that the intrinsic density has two peaks, with the neck
between them having only one-third the peak density; we regard
these as two $\alpha$'s.

Figure~\ref{fig:be4-con} shows the corresponding calculation for
the $J$~=~4$^+$, $M_J$~=~4, state of $^8$Be.  In this case the
the laboratory density does not have to be spherically symmetric
and there is evidence of two $\alpha$'s rotating around the
${\bf z}$ axis in the equatorial plane.  
The projection to the intrinsic frame rotates these to the 
${\bf z}^\prime$ axis and results in an intrinsic density that
is insignificantly different from that obtained for the ground state.
A calculation for the $J$~=~2$^+$ state also produces the same
intrinsic density.

These results, obtained for the VMC wave functions, suggest that
the 0$^+$, 2$^+$, and 4$^+$ wave functions for $^8$Be have the
structure of a deformed rotor
consisting of two $\alpha$'s.  The structure is not
manifest in the shell model part of the VMC 
wave functions; it is induced by the correlations.  The optimum 
spatial correlations between pairs of s-shell and p-shell nucleons 
are similar, but those between one s-shell and one p-shell nucleon 
are different.  Calculations using GFMC configurations
give very similar results except that, especially for the $J$~=~4$^+$
state, the rms radius is still growing at the end of the GFMC propagation.

If the $0^+$, $2^+$, and $4^+$ states are generated by rotations
of a common deformed structure, then their electromagnetic moments
and transition strengths should all be related to the intrinsic
moments.  Table \ref{table:be2} shows the VMC computed values
of the quadrupole and hexadecapole moments, and of the B(E2) and
B(E4) strengths for these states.  In the rotational model these
are related by simple Clebsch Gordon coefficients
to the intrinsic quadrupole, $Q_0$, or $M_4$ moment
depending on the multipolarity, $\lambda$.  
These moments are defined as
\begin{equation}
Q_0 = \sqrt{\case{16\pi}{5}}
\int d {\bf r}^\prime \rho({\bf r}^\prime) 
{{\bf r}^\prime}^2 Y_{20}(\hat{\bf r}^\prime) \ , \\
\end{equation}
and
\begin{equation}
M_4 = 
\int d {\bf r}^\prime \rho({\bf r}^\prime) 
{{\bf r}^\prime}^4 Y_{40}(\hat{\bf r}^\prime) \ , \\
\end{equation}
where $\rho$ is the point proton density and ${\bf r}^\prime$ refers to the
intrinsic (body-fixed) frame.
The last column gives
the extracted values of these intrinsic moments; we see that,
with the exception $M_4$ for the $J = 2^+$ state, these are remarkably
constant.  

We can also compute values of $Q_0$ and $M_4$ by integrating over the
projected body-fixed densities.  These values might be too large
because, as described above, the projection method can introduce
an excessive deformation in the intrinsic shape.  The values
for the $0^+$, $2^+$, and $4^+$ states are $Q_0$ = 26.2, 27.9, and
26.7, and $M_4$ = 55, 62, 64; the $Q_0$ are in good agreement with
the values given in the Table, which were obtained from the spectroscopic
quadrupole moments.
We note that the ratios of these $Q_0$ and $M_4$ values are in reasonable
agreement with the ratio obtained for a diatomic model of $^8$Be
assuming two point $\alpha$'s separated by 4 fm.

We can attempt to project out an intrinsic structure of the
$T$~=~0, $J$~=~1$^+$ and $J$~=~3$^+$, states of $^8$Be.  In this
case we also find two peaks but of somewhat smaller density than for the
0$^+$, 2$^+$, and 4$^+$ states.  Clearly these peaks cannot be
due to two $\alpha$'s rotating around each other.  They presumably 
reflect occurrence of $\alpha+t+p$ and $\alpha+^3$He$+n$ structures 
in these states.

\section{CONCLUSIONS}

We have made quantum Monte Carlo calculations for the ground states and 
low-lying excitations of $A=8$ nuclei interacting by realistic two- and 
three-nucleon potentials.
These calculations have been made practical by the development of a 
constrained-path algorithm for the complex spatial and 
spin-isospin wave functions needed to describe these nuclei.
This algorithm greatly reduces the ``fermion-sign problem'' that quantum
Monte Carlo methods are subject to, allowing us to obtain binding energies 
for a given Hamiltonian that are accurate to 1 to 2\%.

The Hamiltonian we have used, consisting of the Argonne $v_{18}$ two-nucleon
and Urbana IX three-nucleon potentials, gets the general features of the
light p-shell nuclei fairly well, including the bulk of the experimentally
observed binding and the correct ordering and approximate spacing of the 
excitation spectrum.
However, with accurate calculations, we can identify specific
failings of this Hamiltonian, including a few percent underbinding in $N=Z$
nuclei, which gets progressively worse as the neutron-proton asymmetry 
increases, and spin-orbit splittings in the excitation spectrum that are 
too small.
Nevertheless, we can see that the energy differences with experiment are 
much smaller than the magnitude of the short-range (and least-well-known)
part of our three-nucleon interaction, so it is plausible that new
three-nucleon force models may be able to reproduce the ground and
excited states near the 1\% level.

We have also been able to study energy differences between isobaric analog
states and the isospin-mixing matrix elements in $^8$Be.
While final conclusions should be reserved until the bulk energies are
corrected by improved three-nucleon potentials, we can see that the
charge-dependent and charge-symmetry-breaking components of the 
Argonne $v_{18}$ potential are making significant contributions that
improve the agreement with experiment.

Finally, we have also studied the moments, densities, momentum distributions,
and intrinsic shapes of these nuclei.
It appears that the observed static moments 
may be understood with our present microscopic model
after we include meson-exchange current
contributions. 
We also see evidence for strong clustering in the light p-shell
nuclei, particularly the 2$\alpha$ character in $^8$Be.
This clustering is built into the variational wave functions by the strong pair 
correlations, which depend upon the nucleon orbits 
and seem to be preserved in the GFMC propagation.

With the present method, and the continuing rapid increases in computational
power that massively parallel machines are bringing, we are confident 
that we will be able to extend our calculations to the $A=9,10$ nuclei
in the near future, and to $^{12}$C in a few year's time. 

\acknowledgements
The authors thank Dr. Dieter Kurath for many useful suggestions.  The
many-body calculations were made possible by generous grants of time on
the IBM SP and SGI Origin 2000 of the Mathematics and Computer Science
Division, Argonne National Laboratory.  
The GFMC calculations of $^8$Li excited states were made with early-user time on
the IBM SP at the National Energy Research Scientific Center.  
The work of SCP and RBW is
supported by the U. S. Department of Energy, Nuclear Physics Division,
under contract No. W-31-109-ENG-38, that of JC by the U.S. Department of
Energy under contract W-7405-ENG-36, and that of VRP by the U.S. National 
Science Foundation via grant PHY98-00978.

\narrowtext

\begin{table}
\caption{Values of variational parameters in p-shell nuclei; 
in the expression for $V_p$, $L$ is the total orbital angular momentum.}
\begin{tabular}{ldddd}
  $^{2S+1}L[n]$ & $^1$L[4]   & $^{1,3}$L[31] & $^{1,3}$L[22] & $^3$L[211]  \\
\tableline
  $c_{sp}$      &    0.85    &    0.85    &    0.85    &    0.85    \\
  $d_{sp}$ (fm) &    3.2     &    3.2     &    3.2     &    3.2     \\
  $c^{[n]}_{pp}$&    0.1     &    0.2     &    0.3     &    0.4     \\
  $d_{pp}$ (fm) &    3.2     &    3.2     &    3.2     &    3.2     \\
  $V_p$ (MeV)   & --20.0+$L$ & --20.0+$L$ & --20.0+$L$ & --20.0+$L$ \\
  $R_p$ (fm)    &    4.0     &    4.0     &    4.0     &    4.0     \\
  $a_p$ (fm)    &    1.5     &    1.5     &    1.5     &    1.5     \\
\end{tabular}
\label{table:params8}
\end{table}

\begin{table}
\caption{$\beta_{LS[n]}$ components for $^8$He states, listed in order of
increasing excitation.}
\begin{tabular}{cddd}
$(J^{\pi};T)$ & $^1$S[22] & $^1$D[22] & $^3$P[211] \\
\tableline
$(0^+;2)$     &   0.842   &           & --0.539    \\
$(2^+;2)$     &           &   0.958   & --0.287    \\
$(1^+;2)$     &           &           &   1        \\
$(0^+;2)$     &   0.546   &           &   0.838    \\
$(2^+;2)$     &           &   0.294   &   0.956    \\
\end{tabular}
\label{table:beta82}
\end{table}

\begin{table}
\caption{$\beta_{LS[n]}$ components for $^8$Li states, listed in order of
increasing excitation.}
\begin{tabular}{cdddddddd}
$(J^{\pi};T)$ & $^3$P[31] & $^3$D[31] & $^3$F[31] & $^1$P[31]
              & $^1$D[31] & $^1$F[31] & $^3$S[22] & $^3$D[22] \\ 
\tableline
$(2^+;1)$     &   0.936   & --0.289   & --0.121   &
              &   0.118   &           &           & --0.104   \\
$(1^+;1)$     &   0.740   & --0.334   &           &   0.576
              &           &           & --0.067   & --0.068   \\
$(0^+;1)$     &   1.      &           &           &
              &           &           &           &           \\
$(3^+;1)$     &           &   0.912   &   0.275   &
              &           & --0.296   &           & --0.067   \\
$(2^+;1)$     &   0.343   &   0.854   &   0.386   &
              &   0.046   &           &           &   0.047   \\
$(1^+;1)$     &   0.644   &   0.410   &           & --0.618
              &           &           & --0.162   & --0.091   \\
$(1^+;1)$     & --0.036   &   0.812   &           &   0.531
              &           &           & --0.229   &   0.063   \\
$(4^+;1)$     &           &           &   1.      &
              &           &           &           &           \\
$(2^+;1)$     & --0.111   & --0.260   &   0.550   &
              &   0.780   &           &           & --0.101   \\
$(3^+;1)$     &           & --0.269   &   0.922   &
              &           &   0.037   &           & --0.275   \\
\end{tabular}
\label{table:beta81}
\end{table}

\begin{table}
\caption{$\beta_{LS[n]}$ components for $^8$Be states, listed in order of
increasing excitation.}
\begin{tabular}{cdddddd}
$(J^{\pi};T)$ & $^1$S[4] & $^1$D[4] & $^1$G[4]  & $^3$P[31]
                                    & $^3$D[31] & $^3$F[31] \\
\tableline
$(0^+;0)$     &   0.997  &          &           & --0.076
                                    &           &           \\
$(2^+;0)$     &          &   0.999  &           &   0.033
                                    &   0.037   & --0.018   \\
$(4^+;0)$     &          &          &   0.997   &        
                                    &           &   0.079   \\
$(2^+;0)$     &          & --0.019  &           &   0.948
                                    & --0.314   &   0.045   \\
$(1^+;0)$     &          &          &           &   0.861
                                    &   0.508   &           \\
$(1^+;0)$     &          &          &           & --0.526
                                    &   0.850   &           \\
$(3^+;0)$     &          &          &           &        
                                    &   0.891   &   0.453   \\
$(4^+;0)$     &          &          & --0.089   &        
                                    &           &   0.996   \\
$(2^+;0)$     &          & --0.038  &           &   0.270
                                    &   0.844   &   0.463   \\
$(3^+;0)$     &          &          &           &        
                                    & --0.457   &   0.889   \\
$(0^+;0)$     &   0.098  &          &           &   0.995
                                    &           &           \\
\end{tabular}
\label{table:beta80}
\end{table}

\begin{table}
\caption{$^6$Li(gs) results (in MeV) using different three-body propagators.}
\begin{tabular}{lddd}
Propagator              & $\langle V^{2\pi,A}\rangle $ & 
$\langle V^{2\pi,C}\rangle $ & $ \langle H \rangle $ \\
\hline
$G$                     & --8.6(2)        & --5.2(1)        &   --31.25(12) \\
$\tilde{G}, n=1$        & --8.5(2)        & --5.2(1)        &   --31.15(11) \\
$\tilde{G}, n=4$        & --8.9(2)        & --5.3(1)        &   --31.23(7) \\
$\tilde{G}, n=\infty$   & --8.8(2)        & --5.2(1)        &   --31.25(12) \\
just two-body           & --5.8(2)        & --3.9(1)        &   --29.86(23) \\
\end{tabular}
\label{table:newprop}
\end{table}

\begin{table}
\caption{Scaling of calculation with system size}
\begin{tabular}{cccdd}
         & $A$ & $P=A(A-1)/2$ & $2^A\times I(A,T)$ & $\prod (\times^8$Be) \\
\hline
   $^4$He  &  4  &   6 &    16$\times$2     &   0.001     \\
   $^5$He  &  5  &  10 &    32$\times$5     &   0.010     \\
   $^6$Li  &  6  &  15 &    64$\times$5     &   0.036     \\
   $^7$Li  &  7  &  21 &   128$\times$14    &   0.33      \\
   $^8$Be  &  8  &  28 &   256$\times$14    &   1.        \\
   $^8$He  &  8  &  28 &   256$\times$20    &   1.4       \\
   $^8$Li  &  8  &  28 &   256$\times$28    &   2.        \\
   $^9$Be  &  9  &  36 &   512$\times$42    &   8.7       \\
 $^{10}$B  & 10  &  45 &  1024$\times$42    &  24.        \\
 $^{11}$B  & 11  &  55 &  2048$\times$132   & 200.        \\
 $^{12}$C  & 12  &  66 &  4096$\times$132   & 530.        \\
\hline
   $^8$n   &  8  &  28 &   256$\times$1     &   0.071     \\
 $^{14}$n  & 14  &  91 & 16384$\times$1     &  26.        \\
\end{tabular}
\label{table:scaling}
\end{table}

\begin{table}
\caption{Experimental and quantum Monte Carlo energies of $A$=2--8 nuclear
ground states in MeV.}
\begin{tabular}{cdddd}
$^AZ(J^{\pi};T)$ & ~~~VMC ($\Psi_T$) & ~VMC ($\Psi_V$) & GFMC  & Expt  \\
\hline
$^2$H$(1^+;0)$      &  --2.2248(5)&             &              &  --2.2246 \\
$^3$H$(\case{1}{2}^+;\case{1}{2})$
                    &  --8.15(1)  &  --8.32(1)  &   --8.47(1)  &  --8.48  \\
$^4$He$(0^+;0)$     & --26.97(3)  & --27.78(3)  &  --28.34(4)  & --28.30  \\
$^6$He$(0^+;1)$     & --23.64(7)  & --24.87(7)  &  --28.11(9)  & --29.27  \\
$^6$Li$(1^+;0)$     & --27.10(7)  & --27.83(5)  &  --31.15(11) & --31.99  \\
$^7$He$(\case{3}{2}^-;\case{3}{2})$
                    & --18.05(11) & --19.75(12) &  --25.79(16) & --28.82  \\
$^7$Li$(\case{3}{2}^-;\case{1}{2})$
                    & --31.92(11) & --33.04(7)  &  --37.78(14) & --39.24  \\
$^8$He$(0^+;2)$     & --17.98(8)  & --19.31(12) &  --27.16(16) & --31.41  \\
$^8$Li$(2^+;1)$     & --28.00(14) & --29.76(13) &  --38.01(19) & --41.28  \\
$^8$Be$(0^+;0)$     & --45.47(16) & --46.79(19) &  --54.44(19) & --56.50  \\
\end{tabular}
\label{table:energy}
\end{table}

\begin{table}
\caption{VMC, GFMC, and experimental excitation energies (adjusted to their
respective ground states) in MeV.  Numbers in parentheses are Monte Carlo
error estimates for theory, and uncertainties in the energy for experiment.
We also give experimental widths where known in keV (except where otherwise
noted).  The $^8$Be states marked with an * are isospin-mixed.}
\begin{tabular}{cdddd}
$^AZ(J^{\pi};T)$ & ~~~~~VMC   &  ~~~GFMC  &   Expt &  $\Gamma$   \\
                 & ~~~~~(MeV) &  ~~~(MeV) &  (MeV) &  (keV)      \\
\hline
$^8$He$(2^+;2)$  &  2.22(16) &  3.01(23) &   3.59  &             \\
$^8$He$(1^+;2)$  &  3.32(17) &  4.42(25) &         &             \\
$^8$He$(0^+;2)$  &  4.72(18) &           &         &             \\
$^8$He$(2^+;2)$  &  5.01(17) &           &         &             \\
\hline
$^8$Li$(1^+;1)$  &  1.34(18) &  0.57(24) &   0.98  & 12(4) fs    \\
$^8$Li$(0^+;1)$  &  2.83(19) &  1.91(29) &         &             \\
$^8$Li$(3^+;1)$  &  3.35(18) &  2.66(28) &   2.26  & 33(6)       \\
$^8$Li$(2^+;1)$  &  3.86(18) &           &         &             \\
$^8$Li$(1^+;1)$  &  4.22(19) &           &   3.21  & $\approx$ 1000  \\
$^8$Li$(1^+;1)$  &  5.32(19) &           &   5.4   & $\approx$ 650   \\
$^8$Li$(4^+;1)$  &  6.08(18) &  6.27(31) &   6.53  & 35(15)      \\
$^8$Li$(2^+;1)$  &  6.20(18) &           &         &             \\
$^8$Li$(3^+;1)$  &  7.31(18) &           &   6.1   & $\approx$ 1000  \\
$^8$Li$(0^+;2)$  & 11.24(18) & 12.10(25) &  10.82  & $<$ 12        \\
\hline
$^8$Be$(2^+;0)$  &  2.39(25) &  2.91(25) &   3.04(3) & 1500(20)  \\
$^8$Be$(4^+;0)$  &  9.95(24) &  9.58(27) &  11.4(3)  & $\approx$ 3500  \\
$^8$Be$(2^+;1)$  & 18.60(23) & 18.02(27) &  16.63*   &  108(1)  \\
$^8$Be$(2^+;0)$  & 20.29(24) &           &  16.92*   &   74(1)  \\
$^8$Be$(1^+;1)$  & 19.89(23) &           &  17.64    &   11(1)  \\
$^8$Be$(1^+;0)$  & 20.03(23) & 18.09(33) &  18.15    &  138(6)  \\
$^8$Be$(1^+;0)$  & 21.73(24) &           &           &          \\
$^8$Be$(3^+;1)$  & 21.77(22) &           &  19.07    &  270(20)  \\
$^8$Be$(3^+;0)$  & 21.85(24) & 19.53(33) &  19.24    &  230(30)  \\
$^8$Be$(4^+;0)$  & 25.85(22) &           &  19.86    &  700(100)  \\
$^8$Be$(2^+;0)$  & 23.61(23) &           &  20.1     & $\approx$ 1100  \\
$^8$Be$(3^+;0)$  & 25.53(23) &           &           &           \\
$^8$Be$(0^+;0)$  & 28.62(26) &           &  20.2     & $<$ 1000  \\
$^8$Be$(0^+;2)$  & 29.55(22) & 29.93(25) &  27.49    &   5.5(2.0) \\
\end{tabular}
\label{table:excited}
\end{table}

\begin{table}
\caption{Kinetic and potential energy contributions to GFMC energies in MeV.}
\begin{tabular}{cddddddd}
$^AZ(J^{\pi};T)$ &                $K$ & $v_{ij}$ & $V_{ijk}$ & $v^{\gamma}_{ij}$ & $v^{\pi}_{ij}$ & $V^{2\pi}_{ijk}$ & $V^R_{ijk}$ \\
\hline
$^2$H$(1^+;0)$ &                       19.81      &  $-$22.04        &            & 0.018   &  $-$21.28     &        \\
$^3$H$(\case{1}{2}^+;\case{1}{2})$ &   51.2(3)    &  $-$58.9(3)      &  $-$1.2(0) & 0.04(0) &  $-$45.0(2)   &  $-$2.2(0) &  1.0(0) \\
$^4$He$(0^+;0)$ &                     110.7(7)    & $-$135.3(7)      &  $-$6.3(1) & 0.86(0) & $-$102.4(5)   & $-$11.5(1) &  5.2(1) \\
$^6$He$(0^+;1)$ &                     138.(1)     & $-$164.(1)       &  $-$7.0(1) & 0.86(0) & $-$119.(1)    & $-$13.1(2) &  6.1(2) \\
$^6$Li$(1^+;0)$ &                     153.(1)     & $-$183.(1)       &  $-$7.3(2) & 1.68(1) & $-$143.(1)    & $-$13.6(2) &  6.3(2) \\
$^7$He$(\case{3}{2}^-;\case{3}{2})$ & 156.(2)     & $-$183.(2)       &  $-$8.1(2) & 0.87(0) & $-$131.(1)    & $-$15.2(4) &  7.2(3) \\
$^7$Li$(\case{3}{2}^-;\case{1}{2})$ & 190.(2)     & $-$227.(2)       &  $-$9.1(2) & 1.77(1) & $-$172.(1)    & $-$17.9(3) &  8.8(2) \\
$^8$He$(0^+;2)$ &                     166.(1)     & $-$195.(1)       &  $-$8.0(2) & 0.87(0) & $-$135.(1)    & $-$15.8(2) &  7.8(2) \\
$^8$Li$(2^+;1)$ &                     217.(2)     & $-$257.(2)       & $-$10.2(2) & 1.86(1) & $-$192.(1)    & $-$20.8(4) & 10.6(3) \\
$^8$Be$(0^+;0)$ &                     248.(2)     & $-$301.(2)       & $-$14.9(3) & 3.27(1) & $-$227.(1)    & $-$27.1(4) & 12.3(3) \\
\end{tabular}
\label{table:gfmc_terms}
\end{table}

\begin{table}
\caption{VMC and GFMC isovector and isotensor energy coefficients $a^{(n)}_{A,T}$ in keV.}
\begin{tabular}{cccccccccc}
$a^{(n)}_{A,T}$
& \multicolumn{2}{c}{$K^{CSB}$} & \multicolumn{2}{c}{$v^\gamma$} 
& \multicolumn{2}{c}{$v^{CSB}+v^{CD}$} 
& \multicolumn{3}{c}{Total} \\
              & VMC & GFMC & VMC & GFMC & VMC & GFMC &  VMC    &  GFMC & Expt.  \\
\hline
$a^{(1)}_{8,1}$ & 24 & 20 & 1689 & 1586 &  ~66 & 63 & 1779(4)  & 1669(7)  & 1770 \\
$a^{(2)}_{8,1}$ & ~0 & ~0 & ~155 & ~132 & --11 &~~3 & ~144(4)  & ~135(7)  &~~145 \\
$a^{(1)}_{8,2}$ & 17 & 15 & 1478 & 1423 &  ~40 & 47 & 1535(3)  & 1485(5)  & 1659 \\
$a^{(2)}_{8,2}$ & ~0 & ~0 & ~140 & ~129 &  ~26 & 29 & ~166(4)  & ~158(4)  &~~153 \\
\end{tabular}
\label{table:analog}
\end{table}

\begin{table}
\caption{VMC isospin-mixing matrix elements for $^8$Be in keV}
\begin{tabular}{lcccccc}
$J^{\pi}$ & $v^{\rm Coul}$ & $v^{\rm MM}$ & $v^{\rm CSB}$ & $K^{\rm CSB}$ &  
$E_{01}$ & Expt. \\
\hline
2$^+$     &   63   &   18   &   26   &   2   &  109(4)  &   149   \\
1$^+$     &   40   &  --1   &   15   &   1   &  ~55(2)  &   120   \\
3$^+$     &   33   &   15   &   13   &   1   &  ~62(2)  &   ~63   \\
\end{tabular}
\label{table:mixing}
\end{table}

\begin{table}
\caption{VMC and GFMC values for point proton rms radii (in fm), for 
magnetic moments (in $\mu_N$), and quadrupole moments (in fm$^2$) in 
impulse approximation.
In the GFMC calculations, the rms radii of the two excited states of $^8$Be 
are steadily growing with propagation time.}
\begin{tabular}{lccccccccc}
& \multicolumn{2}{c}{$\langle r^2_p \rangle^{1/2}$}
& \multicolumn{3}{c}{$\mu$}
& \multicolumn{3}{c}{$Q$} 
& \multicolumn{1}{c}{$\langle r^4_p Y_{40} \rangle$} \\
                &  VMC    &    GFMC     &   VMC   &    GFMC & Expt. &    VMC   &       GFMC     &  Expt.   & VMC \\
\hline                                                                                         
$^8$He$(0^+;2)$ & 1.91(1) &  ~1.98(1)   &         &         &       &          &                &          & \\
$^8$Li$(2^+;1)$ & 2.21(2) &  ~2.18(2)   & 1.12(1) & 1.3(1)~ & 1.653 &  ~3.2(1) &  ~~~3.6(2)     & 3.19(7)~ & 0.1(2) \\
$^8$Be$(0^+;0)$ & 2.25(1) &  ~2.48(1)   &         &         &       &          &                &          & \\
$^8$B$(2^+;1)$  & 2.47(3) &  ~2.57(1)   & 1.46(1) & 1.4(2)~~& 1.036 &  ~5.5(2) &  ~~~7.9(4)     & 6.83(21) & 0.7(8) \\
$^8$C$(0^+;2)$  & 2.90(4) &  ~3.17(1)   &         &         &       &          &                &          & \\
\end{tabular}
\label{table:radii}
\end{table}

\begin{table}
\caption{VMC calculation of moments and transition strengths for different
multipolarities, $\lambda$, in the
rotational states of $^8$Be; the units are fm and the proton charge.}
\begin{tabular}{ccdd}
Observable                                  & $\lambda$ &   Value   & $Q_0$ or $M_4$ \\
\hline
Q($2^+$)   &                                       2 &  $-$7.61(8)  & 26.6(3)   \\
Q($4^+$)   &                                       2 &  $-$9.76(10) & 26.8(3)   \\
B(E2, $2^+ \rightarrow 0^+$) &                     2 &    14.8(4)   & 27.3(4)   \\
B(E2, $4^+ \rightarrow 2^+$) &                     2 &    18.2(4)   & 25.3(3)   \\
\\
$\langle 2^+,2 | r^4 Y_{4,0} | 2^+,2 \rangle$  &   4 &     2.5(2)   & 53.(4) \\
$\langle 4^+,4 | r^4 Y_{4,0} | 4^+,4 \rangle$  &   4 &     4.3(2)   & 34.(2) \\
B(E4, $4^+ \rightarrow 0^+$) &                     4 &   167.(16)   & 39.(2) \\
\end{tabular}
\label{table:be2}
\end{table}

\begin{figure}
\caption{The experimental spectrum for $A=8$ nuclei. All narrow low-lying 
states are shown, as well as the wide rotational states in $^8$Be.  
Higher-lying and negative parity states in $^8$Be and some wider states
in $^8$Li are not shown.}
\label{fig:expt8}
\end{figure}

\begin{figure}
\caption{Constrained and unconstrained propagation tests for $^6$Li(gs).  See the text for a description of the symbols.}
\label{fig:e-of-tau_6Li-test}
\end{figure}

\begin{figure}
\caption{Constrained and unconstrained propagation tests for an eight-neutron
drop.  See the text for a description of the symbols.}
\label{fig:e-of-tau_8n-test}
\end{figure}

\begin{figure}
\caption{Constrained and unconstrained propagation tests for $^8$He(gs).  See the text for a description of the symbols.}
\label{fig:e-of-tau_8He-test}
\end{figure}

\begin{figure}
\caption{Constrained propagation to large $\tau$ for states of $^6$Li.}
\label{fig:e-of-tau_6Li-largetau}
\end{figure}

\begin{figure}
\caption{GFMC $E(\tau)$ for states of $^8$Be.  The solid lines show
the averages of the last 7 values; the dashed lines show the corresponding
statistical errors.}
\label{fig:e-of-tau_8Be}
\end{figure}

\begin{figure}
\caption{GFMC $E(\tau)$ for states of $^8$Li.}
\label{fig:e-of-tau_8Li}
\end{figure}

\begin{figure}
\caption{GFMC $E(\tau)$ for states of $^8$He.}
\label{fig:e-of-tau_8He}
\end{figure}

\begin{figure}
\caption{VMC, GFMC, and experimental energies of nuclear states for
$4 \leq A \leq 8$.  The light shading shows the Monte Carlo statistical
errors or experimental uncertainties.}
\label{fig:energies}
\end{figure}

\begin{figure}
\caption{VMC, GFMC, and experimental excitation energies of nuclear states for
$6 \leq A \leq 8$}
\label{fig:excitations}
\end{figure}

\begin{figure}
\caption{The neutron and proton densities in $^8$He, $^8$Li, and $^8$Be.}
\label{fig:8bodyrho1}
\end{figure}

\begin{figure}
\caption{The neutron and proton densities in $^4$He, $^6$He, and $^8$He,
shown on a logarithmic scale.}
\label{fig:herho1}
\end{figure}

\begin{figure}
\caption{The proton-proton densities in $^4$He, $^6$He, and $^8$He.}
\label{fig:herho2}
\end{figure}

\begin{figure}
\caption{The neutron and proton momentum distributions in $^8$He, $^8$Li,
and $^8$Be.}
\label{fig:8bodyrhok}
\end{figure}

\begin{figure}
\caption{Contours of constant density, plotted in cylindrical coordinates, for $^8$Be(0$^+$).
The left side is in the ``laboratory'' frame while the right side is in the intrinsic frame.}
\label{fig:be0-con}
\end{figure}

\begin{figure}
\caption{Contours of constant density for $^8$Be(4$^+$).}
\label{fig:be4-con}
\end{figure}

\end{document}